\documentclass[prd,a4paper,twocolumn,nofootinbib,aps,10pt,preprintnumbers,longbibliography]{revtex4-2}

\usepackage[colorlinks=true,linkcolor=black,citecolor=blue,urlcolor=blue, pdfborder={0 0 0}]{hyperref}
 \usepackage{amssymb,amsmath,amsthm,cancel,graphicx}
\usepackage{mathtools}
\usepackage[utf8]{inputenc}
\usepackage{color}
\usepackage[table,xcdraw,dvipsnames]{xcolor}
\usepackage{multirow}
\usepackage[normalem]{ulem}    
\usepackage{bm} 
\usepackage{slashed} 
\usepackage{booktabs}
\usepackage{placeins}
\usepackage[table,xcdraw,dvipsnames]{xcolor}
\hypersetup{
	bookmarks=true,         
	unicode=true,          
	pdftoolbar=true,        
	pdfmenubar=true,        
	pdffitwindow=false,     
	pdfstartview={FitH},    
	pdfauthor={Author},     
	pdfsubject={Subject},   
	pdfcreator={Creator},   
	pdfproducer={Producer}, 
	pdfkeywords={keyword1} {key2} {key3}, 
	pdfnewwindow=true,      
	colorlinks=true,       
	linkcolor=black,          
	citecolor=black,        
	filecolor=black,      
	urlcolor=black           
}
\usepackage{physics} 
\usepackage{soul}

\def\eq#1{{Eq.~(\ref{#1})}}

\newcommand{\be}{\begin{equation}}
\newcommand{\ee}{\end{equation}}
\newcommand{\bea}{\begin{eqnarray}}
\newcommand{\eea}{\end{eqnarray}}

\newcommand{\D}{\scriptscriptstyle{D}}

\newcommand{\gammaD}{\gamma_{\scriptscriptstyle{D}}}

\newcommand{\rDP}{r_{\!\D}}
\newcommand{\mDP}{m_{\gamma_{\!\lower0.4ex\hbox{$\scriptscriptstyle D$}}}}
\newcommand{\mpi}{m_{\pi}}
\newcommand{\mK}{m_{K}}

\newcommand{\rp}{r_{\!\pi}}

\newcommand{\BR}{{\rm BR}}
\newcommand{\Lambdasd}{\Lambda_{sd}}
\newcommand{\alphaD}{\alpha_{\D}}
\newcommand{\FA}{F_{\scriptscriptstyle{A}}}
\newcommand{\FB}{F_{\scriptscriptstyle{B}}}
\newcommand{\FC}{F_{\scriptscriptstyle{C}}}

\newcommand{\fAA}{f_{\scriptscriptstyle{AA}}}
\newcommand{\fBB}{f_{\scriptscriptstyle{BB}}}
\newcommand{\fCC}{f_{\scriptscriptstyle{CC}}}
\newcommand{\fAB}{f_{\scriptscriptstyle{AB}}}
\newcommand{\fAC}{f_{\scriptscriptstyle{AC}}}
\newcommand{\fBC}{f_{\scriptscriptstyle{BC}}}
\newcommand{\DiscB}{{\rm DiscB}}
\newcommand{\KKbar}{\ensuremath{K^0\!\scalebox{0.75}[1]{$-$} \bar K^0}}

\newcommand*\xbar[1]{%
  \hbox{\;%
    \vbox{%
      \hrule height 0.5pt 
      \kern0.5ex
      \hbox{%
        \kern-0.25em
        \ensuremath{#1}%
        \kern-0.07em
      }%
    }%
  }%
} 

\definecolor{KBFIred}{RGB}{163,35,47}

\begin{document}
	
\title{\Large \textbf{\color{KBFIred}{
      Sensitivity targets for dark photon searches\\ in charged kaon decays
}}}

\vspace{1cm}
\author{\bf M. Fabbrichesi$^{{a}}$}
\author{\bf E. Gabrielli$^{{b,a,c}}$}
\author{\bf K. M\"u\"ursepp$^{{c,d}}$}
\vspace{0.5cm}

\affiliation{$^{a}$ INFN, Sezione di Trieste, Via Valerio 2, I-34127 Trieste, Italy}
\affiliation{$^{b}$ Physics Department, University of Trieste, Strada Costiera 11, I-34151 Trieste, Italy}
\affiliation{$^{c}$ Laboratory of High-Energy and Computational Physics, NICPB, R\"avala 10,  10143 Tallinn, Estonia}
\affiliation{$^{d}$ INFN, Laboratori Nazionali di Frascati, C.P. 13, 100044 Frascati, Italy  }

\begin{abstract}
We investigate the prospects for detecting invisible dark photons in charged
kaon decays through the two-pion channel
$K^+\to\pi^+\pi^0\,\gammaD$ and the one-pion radiative decay
$K^+\to\pi^+\gamma\,\gammaD$, where $\gamma$ and $\gammaD$ denote the
ordinary photon and dark photon, respectively.
The former decay was previously proposed as a golden channel for massless dark photon searches at NA62, while here we introduce the radiative process as a complementary probe and extend both analyses to massive dark photons.
The three-body modes are particularly relevant at low dark photon masses, where
the two-body decay $K^+\to\pi^+\,\gammaD$ becomes strongly suppressed and
vanishes in the massless limit. 
For massless dark photons, renormalizable interactions induced by kinetic mixing are absent, and the dominant contribution arises from the effective non-renormalizable flavor-changing quark interaction $s\to d\,\gammaD$ mediated by a dimension-five magnetic-dipole operator. This is also true for massive dark photons. Their kinetic mixing contributes negligibly
to these channels, after considering the existing bounds on the mixing parameter. Only the same flavor changing magnetic-dipole operator as in the massless case is relevant. Using the chiral quark model, we estimate the relevant hadronic matrix elements of the dipole operator and account for constraints from $K^+\to\pi^+\,X$, where $X$ denotes resonant missing energy, $K^+\to\pi^+\nu\bar\nu$, and \KKbar~mixing. Based on these calculations, we determine the allowed branching ratios for these decay modes and the sensitivities required at NA62 and future charged-kaon facilities.
\end{abstract}

		\vspace{1cm}
		\maketitle

\section{Introduction}
\label{sec:intro}
Astrophysical and cosmological observations provide evidence for the existence of dark matter, suggesting the possibility of a dark sector whose particles remain neutral under the known Standard Model (SM) gauge
interactions~\cite{Bertone:2004pz}.
If the dark and visible sectors interacted solely through gravity, there would be little chance of directly probing dark-sector particles in laboratory experiments. Instead, a connection between the two sectors could arise through portal interactions mediated by new fields that communicate with both sectors~\cite{Beacham:2019nyx,Batell:2022dpx}.

Among the simplest and most extensively investigated possibilities is the 
existence of an additional secluded $U(1)_D$ gauge symmetry, with interactions 
in the dark sector mediated by a dark photon $\gammaD$ 
(see, for instance, ~\cite{Essig:2013lka,Alexander:2016aln,Fabbrichesi:2020wbt,Lanfranchi:2020crw, Antel:2023hkf, Goudzovski:2022vbt} for recent reviews). The dark photon, the gauge boson associated with the $U(1)_D$ symmetry, can be either massive or massless, depending on whether this gauge symmetry is broken or remains unbroken, as discussed in \cite{Holdom:1985ag,delAguila:1995rb}.
\cite{Holdom:1985ag,delAguila:1995rb}.

Most experimental searches have traditionally focused on the massive dark
photon scenario, while the phenomenology and experimental prospects of massless
dark photons have also received increasing attention in recent years
\cite{Gabrielli:2016cut,Gabrielli:2016vbb,Fabbrichesi:2017vma,Cobal:2020hmk,Biswas:2017lyg,Fabbrichesi:2017zsc,Casarsa:2021rud,Gabrielli:2019sjg,Gabrielli:2021lkw,Gabrielli:2024wys}, including searches exploiting
Higgs-boson production and decays at the LHC
\cite{Gabrielli:2014oya,Biswas:2015sha,Biswas:2016jsh,
ATLAS:2026bqr,CMS:2020krr,Biswas:2022tcw}.
The possible role of massless dark photons in dark-matter dynamics has also
been extensively discussed
\cite{Foot:2004pa,Ackerman:2008kmp,Fan:2013tia,Foot:2014uba,
  Heikinheimo:2015kra,Agrawal:2016quu}.

The phenomenology of the two scenarios, as well as the corresponding search
strategies, can be substantially different. A massless dark photon does not
acquire direct renormalizable interactions with SM fields through kinetic
mixing with the ordinary photon. Indeed, in the presence of two unbroken
$U(1)$ gauge symmetries, the kinetic-mixing term can be removed by a suitable
field redefinition, leaving no direct coupling of the massless dark photon to
SM currents. Nevertheless, in the presence of a portal messenger sector, effective interactions between a massless dark photon and ordinary matter are possible.
They arise at low energies through higher-dimensional operators.  Such operators are
typically generated at the loop level by heavy messenger fields charged under
both the SM and the unbroken $U(1)_D$ gauge symmetries, with the Wilson
coefficients suppressed by inverse powers of the heavy new-physics scale.
Their contributions to physical amplitudes, however, can grow with the characteristic energy of the process due to their higher-dimensional nature,
until the cut-off scale of the effective field theory.
Consequently, they naturally satisfy stringent low-energy constraints while becoming increasingly relevant at higher energies.

Among the effective non-renormalizable interactions, dimension-five magnetic-dipole operators provide the leading couplings to SM fermions. Constraints on such interactions have been derived both from laboratory 
experiments and astrophysical observations \cite{Fabbrichesi:2020wbt}. Of particular interest for kaon physics are 
flavor-changing neutral-current (FCNC) transitions of the form 
$q\to q'\, \gammaD$, induced by flavor-changing dark magnetic-dipole operators, 
which can give rise to potentially observable rare decays~\cite{Gabrielli:2016cut}. In 
particular, within this framework, the $s\to d\, \gammaD$ transition provides the 
leading mechanism for massless dark photon production in the kaon decays 
considered here.

The situation is different for a massive dark photon. Once $U(1)_D$ is broken, 
kinetic mixing induces a physical renormalizable coupling of $\gammaD$ to the 
electromagnetic current, proportional to the mixing parameter $\varepsilon$ \cite{Holdom:1985ag,delAguila:1995rb,Fabbrichesi:2020wbt}. This 
interaction constitutes the conventional production mechanism targeted by most 
dark photon searches. Higher-dimensional interactions, including the 
flavor-changing magnetic-dipole operator discussed above, can nevertheless 
also remain present  in the massive case. The two mechanisms can therefore 
provide complementary contributions to dark photon production in rare kaon 
decays.

Dark photon production induced by the FCNC magnetic-dipole interaction through
the three-body decay $K^+\to\pi^+\pi^0\gammaD$ was previously proposed
in Ref.~\cite{Fabbrichesi:2017vma} as a golden channel for massless dark photon searches in charged kaon decays.
Indeed, the corresponding two-body decay $K^+\to\pi^+\gammaD$ is forbidden
in the massless dark photon scenario by angular momentum conservation.

Within a simplified dark-sector model containing scalar messenger fields and
dark fermions~\cite{Gabrielli:2016cut}, it was shown that the FCNC interaction responsible
for the $s\to d\,\gammaD$ transition, and hence for
$K^+\to\pi^+\pi^0\,\gammaD$, can be constrained by
\KKbar~mixing~\cite{Fabbrichesi:2017vma}. The resulting upper bound,
\mbox{${\rm BR}(K^+\to\pi^+\pi^0\,\gammaD)\lesssim 6.9\times10^{-7}$},
lies within the potential sensitivity of NA62, making this channel a
particularly promising probe of massless dark photons in charged kaon decays
\cite{Fabbrichesi:2017vma}.

More recently, it was pointed out that the same FCNC magnetic-dipole operator
can also induce hyperon transitions involving the emission of a dark photon,
such as $\Lambda\to n\,\gammaD$, $\Sigma^+\to p\,\gammaD$,
$\Xi^{(0,-)}\to\Sigma^{(0,-)}\,\gammaD$, and
$\Omega^-\to\Sigma^-\,\gammaD$, providing additional probes of the
$s\to d\,\gammaD$ interaction~\cite{Su:2019ipw}.
Within the ultraviolet (UV) framework of Ref.~\cite{Fabbrichesi:2017vma}, and using the
hadronic inputs adopted in Ref.~\cite{Su:2019ipw},
Ref.~\cite{Su:2020xwt} obtains, for a massless dark photon,
${\rm BR}(K^+\to\pi^+\pi^0\,\gammaD)\lesssim 2.4\times10^{-6}$.
Using the bound on the
effective scale derived in Ref.~\cite{Su:2019ipw}, we obtain instead
${\rm BR}(K^+\to\pi^+\pi^0\,\gammaD)\lesssim 3.9\times10^{-5}$.
We stress that our present calculation reproduces the result obtained
in Ref.~\cite{Fabbrichesi:2017vma}, which is also cited in
Ref.~\cite{Su:2020xwt}. The one order of magnitude discrepancy
can be traced to the different hadronic
matrix elements adopted in Ref.~\cite{Su:2020xwt} for the
$K^+\to\pi^+\pi^0\,\gammaD$ decay amplitude.

These hyperon transitions can also have important astrophysical implications.
In particular, the thermal population of $\Lambda$ hyperons present in the
hot and dense proto-neutron-star core formed during a supernova explosion can
produce dark photons through $\Lambda\to n\,\gammaD$.
Sufficiently weakly interacting dark photons can escape the stellar core,
providing an additional cooling mechanism constrained by the duration of the
neutrino signal observed from SN1987A~\cite{Camalich:2020wac}.
The corresponding limit currently provides the strongest constraint on the
$s\to d\,\gammaD$ magnetic-dipole interaction. Ref.~\cite{Camalich:2020wac}
translated this astrophysical constraint into a bound on
${\rm BR}(K^+\to\pi^+\pi^0\,\gammaD)$ using the kaon-decay result discussed
above. Since our evaluation of the corresponding hadronic matrix element
differs from that adopted there, the same astrophysical bound on the effective
scale leads to a different constraint on the charged-kaon decay rate.
Using our result, consistently with the calculation of
Ref.~\cite{Fabbrichesi:2017vma}, we obtain
\mbox{${\rm BR}(K^+\to\pi^+\pi^0\,\gammaD)\lesssim 1.28\times10^{-9}$}, which is almost three order of magnitude stronger than the constraint from the
$\KKbar$ mixing.

This astrophysical constraint should, however, be interpreted with some caution. Its derivation relies on the properties assumed for the proto-neutron-star core---in particular the density and temperature profile and the nucleon effective mass---which are not directly measured but obtained from numerical simulations, and it further assumes that the emitted dark photons free-stream out of the core without significant reabsorption. Varying these inputs within their present uncertainties has been found to shift the resulting bound by up to an order of 
magnitude~\cite{Rrapaj:2015wgs}. More fundamentally, the robustness of the standard SN1987A cooling argument itself---independent of any dark-photon-specific modeling---has been questioned in~\cite{Bar:2019ifz,Page:2020gsx}, who point out that the observed SN1987A neutrino signal is also compatible with cooling scenarios in which anomalous energy losses play a much larger role than in the fiducial neutrino-driven picture usually assumed. We therefore treat the SN1987A bound as indicative rather than as a firm constraint, and do not use it in deriving the sensitivity targets presented below.

Here, we extend the program of probing dark photons
in charged-kaon
decays in two directions. We first investigate
the one-pion radiative kaon decay $K^+\to\pi^+\gamma \, \gammaD$, where $\gamma$ denotes
the ordinary photon, as a new complementary channel for dark photon searches.
This channel, together with the two-pion decay channel
$K^+\to\pi^+\pi^0\, \gammaD$, is currently under
investigation by the NA62 Collaboration~\cite{NA62:private}.
We then extend the analysis of both channels to the
massive dark photon scenario. For both processes, we estimate the relevant
hadronic matrix elements within the $\chi$QM at the one-loop level.

For massive dark photons, the existing experimental constraints on the mixing parameter $\varepsilon$ (see, for instance, the review \cite{Fabbrichesi:2020wbt}) limit the contributions to both decay modes induced by the conventional kinetic-mixing interaction to be several orders of magnitude beyond the present NA62 sensitivity over most of the relevant parameter space. In contrast, the flavor-changing magnetic-dipole interaction emerges as the leading mechanism for probing dark photons through these channels.

We also take into account the existing constraints on dark photon production through the flavor changing magnetic-dipole operator, for both the massless and massive cases, with a special emphasis on the recent NA62 search for the decay $K^+\to\pi^+X$~\cite{NA62:2025upx}, where $X$ denotes an
invisible particle, and on the recent measurement of
$K^+\to\pi^+\nu\bar\nu$~\cite{NA62:2024pjp,NA62:2026rwr}, which is in good
agreement with the SM prediction~\cite{Buras:2015qea,Buras:2022wpw,
Buchalla:1998ba,Buras:2005gr,Brod:2010hi}.   In addition, recent bounds on the decays of hyperons to hadrons plus invisible states by BESIII~\cite{BESIII:2023utd,BESIII:2026ztz} can also be utilized.
These results provide stringent constraints on the
flavor-changing  $s\to d\, \gammaD$ magnetic-dipole interaction and,
consequently, on the branching ratios ($\BR$s) of the kaon decay modes considered here.
However, we find that these constraints are weaker than one derived from the $\KKbar$ mixing and so we will not consider in the present analysis.
Interestingly, for a massive dark photon the two-body decay
$K^+\to\pi^+\,\gammaD$ becomes increasingly suppressed as $m_{\gammaD}$
decreases, approaching the forbidden massless limit discussed above. The
three-body modes considered here therefore provide complementary probes in
the low-mass region where the sensitivity of the two-body channel deteriorates.

The paper is organized as follows. In Sec.~\ref{sec:amplitudes}, we present the effective interactions and compute the decay amplitudes and rates for the
$K^+\to\pi^+\pi^0\gammaD$, $K^+\to\pi^+\gamma\,\gammaD$, and
$K^+\to\pi^+\gammaD$ processes, including the contribution induced by
kinetic mixing. In Sec.~\ref{sec:KKconstraints}, we derive the constraint on the effective
scale $\Lambdasd$ from $\KKbar$ mixing. In Sec.~\label{sec:SN-caveats}, we discuss the
astrophysical constraints and their limitations. In Sec.~\ref{sec:phenobounds}, we combine
the relevant constraints with the calculated decay rates and discuss
the resulting phenomenological bounds and sensitivity targets.
Our conclusions are presented in Sec.~\ref{sec:conclusions}.
The explicit expressions for
the loop form factors are collected in Appendix.

\section{Amplitudes and decay rates}
\label{sec:amplitudes}

We start from the effective Hamiltonian ${\cal H}^{\Delta S=1}_{\rm eff}$ at
the quark level, responsible for the $\Delta S=1$ FCNC transition
$s\to d\,\gammaD$. We assume this transition to be dominated by a
magnetic-dipole interaction, which can induce the three-body decays
$K^+\to\pi^+\pi^0\,\gammaD$ and
$K^+\to\pi^+\gamma\,\gammaD$, as well as the two body decay $K^+\to\pi^+\gammaD$.
The corresponding effective Hamiltonian can be
parametrized as
\bea
{\cal H}^{\Delta S=1}_{\rm eff}
= \frac{\hat{Q}}{2\Lambdasd}\, ,
\label{Hmdipole}
\eea
where
\bea
\hat{Q}\equiv (\bar{s}\,\sigma^{\mu\nu}d)\,\bar{F}_{\mu\nu}
\label{mdipole}
\eea
is the local
flavor-changing magnetic-dipole operator, with $s$ and $d$ the strange
and down quark fields respectively, $F^{\mu\nu}_{\D}$ denotes the dark-photon
($\gammaD$) field strength, and
$\sigma^{\mu\nu}=\frac{1}{2}[\gamma^\mu,\gamma^\nu]$.
Color and spin contractions are understood.

\begin{figure}[t]
\centering
\includegraphics[width=0.98\columnwidth]{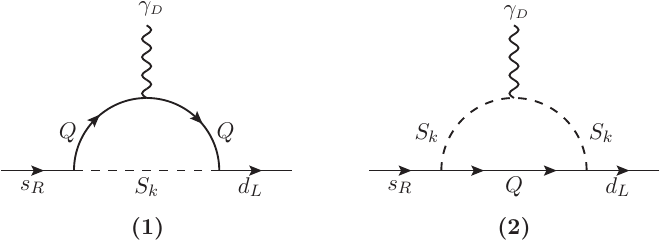}
\caption{\small Vertex diagrams for the generation of the dipole operator in the simplified model of the dark sector (same for the specific model in
\cite{Gabrielli:2016vbb,Gabrielli:2016cut}).}
\label{vertex} 
\end{figure}

As shown in
Refs.~\cite{Fabbrichesi:2017vma,Fabbrichesi:2020wbt,Gabrielli:2016cut},
the magnetic-dipole interaction in \eq{Hmdipole} can be generated at the one-loop level by a dark sector
containing scalar messenger fields and dark fermions running in the loop. See
Fig. \ref{vertex} for illustrative example.

The specific realization adopted to relate the $\Delta S=1$ magnetic-dipole
interaction to the corresponding $\Delta S=2$ contribution to
\KKbar~mixing will be discussed below.

The same UV-complete framework can also generate, for a massive dark
photon, an FCNC vector interaction associated with the effective
dimension-four operator
\bea
\hat{Q}^{V}_{sd}
\equiv
(\bar{s}\gamma_\mu d)\,A_{\D}^{\mu}\, ,
\label{vectoroperator}
\eea
where $A_{\D}^{\mu}$ denotes the dark-photon field. Although
$\hat{Q}^{V}_{sd}$ has canonical dimension four, its Wilson coefficient
is not momentum independent. This can be seen by considering the
corresponding effective $s$-$d$-$\gammaD$ vertex in momentum space,
after separating out the magnetic-dipole contribution. We denote its
remaining vector-current component by $\Gamma_V^\mu(q)$, with $q$ the
momentum carried by the dark photon. Gauge invariance and the associated
Ward identity require this component of the vertex to be transverse and,
up to an overall dimensionless coefficient, its leading
momentum-dependent structure can be parametrized as
\bea
\Gamma^\mu_V(q)
=
\frac{C_V}{\Lambda_V^2}
\left(q^2\gamma^\mu-q^\mu\slashed q\right),
\label{vectorvertex}
\eea
where $\Lambda_V$ denotes the characteristic UV scale associated with
the vector interaction and $C_V$ is a dimensionless coefficient. For
the on-shell decay $s\to d\,\gammaD$, one has $q^2=\mDP^2$ and
$q_\mu\varepsilon_{\D}^{\mu}(q)=0$, where
$\varepsilon_{\D}^{\mu}(q)$ denotes the dark-photon polarization vector.
The corresponding amplitude therefore becomes
\bea
{\cal M}_V(s\to d\,\gammaD)
=
C_V\,\frac{\mDP^2}{\Lambda_V^2}\,
\left(\bar{u}_d\gamma_\mu u_s\right) \,\varepsilon^{\mu}_{{\D}}(q), ,
\label{vectoronshell}
\eea
where $u_s$ and  $u_d$ indicate
the corresponding Dirac spinors in momentum space
for strange and down quarks respectively.
Hence, for an on-shell massive dark photon, the effective Wilson
coefficient multiplying $\hat{Q}^{V}_{sd}$ is suppressed as
$\mDP^2/\Lambda_V^2$ and vanishes in the massless dark photon limit.
The squared vector contribution to the decay rate therefore scales as
$\mDP^4/\Lambda_V^4$, in contrast to the magnetic-dipole contribution,
whose rate scales as $1/\Lambdasd^2$. The interference term is also
parametrically suppressed by the additional factor
$\mDP^2/\Lambda_V^2$ at the amplitude level. Therefore, assuming
$\Lambda_V$ to be of the same order as the UV scale generating the
magnetic-dipole interaction, the vector-current contribution is
negligible compared with the magnetic-dipole one for the dark-photon
masses considered here, and will be neglected in the following.

\subsection{Two-pion $K^+\to \pi^+ \pi^0\, \gammaD$ decay }
\label{sec:Kdp1}

The effective Hamiltonian in \eq{Hmdipole} contributes to the decay process
\bea
K^+(p)\to \pi^+(q_1)\,  \pi^0(q_2)\, \gammaD(k)\, 
\eea
where the $p,q_1,q_2,k$ are the corresponding particles 4-momenta.
The quantum amplitude for this process
\be
\hat{M}\equiv \langle \gammaD\, \pi^0\, \pi^+| {\cal H}^{\Delta S=1}_{\rm eff}| K^+ \rangle
\label{M1}
\ee
has been computed in \cite{Fabbrichesi:2017vma} for the massless dark photon scenario. We extend here the previous results to include the case of a massive dark photon scenario, interacting via the flavor-changing magnetic dipole interaction in \eq{Hmdipole}.
Following the analysis of \cite{Fabbrichesi:2017vma}, we introduce the two dimensionless variable $z_{1,2}$ defined as
\bea
z_{1}=\frac{k\cdot q_1}{m_K^2}\, ~~~~z_{2}=\frac{k\cdot q_2}{m_K^2}\, .
\eea

\begin{figure}[t]
\centering
\includegraphics[width=0.98\columnwidth]{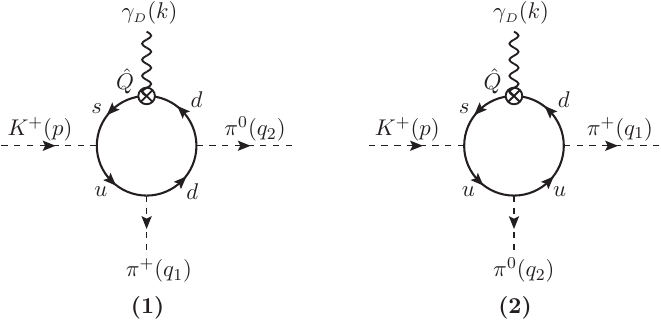}
\caption{\small $\chi$QM diagrams for the process
$K^+\rightarrow\pi^+\,\pi^0\,\gammaD$. The crossed circle stands for
  the insertion of the FCNC magnetic-dipole operator $\hat Q$ in
\eq{mdipole}.}
\label{diagrams1}
\end{figure}

The hadronic matrix element
$\langle \gammaD\,\pi^0\,\pi^+|\hat{Q}|K^+\rangle$
associated with the magnetic-dipole operator in \eq{mdipole} is evaluated
within the $\chi$QM~\cite{Nishijima:1959,Gursey:1959yy,Gursey:1961,Cronin:1967jq,Weinberg:1978kz,Manohar:1983md,Manohar:1984uq}. In this framework,
mesons couple effectively to constituent quarks, allowing the relevant hadronic
matrix elements to be computed from quark-loop diagrams.
The corresponding one-loop Feynman diagrams are shown in Fig.\ref{diagrams1}, where the crossed circle stands for the insertion of the magnetic dipole operator
$\hat{Q}$.

Although the $\chi$QM contains several parameters in general, only the
constituent-quark mass $M$ and the pion decay constant $f$ enter the present
calculation~\cite{Antonelli:1995nv,Fabbrichesi:2017vma}. The $\chi$QM has
been successfully applied to kaon phenomenology~\cite{Antonelli:1995nv}; in
particular, fits to CP-conserving nonleptonic neutral-kaon decay amplitudes
yield
$M\simeq 200\,{\rm MeV}$, with an uncertainty below the $5\%$ level
~\cite{Bertolini:2000dy}. Accordingly, in the evaluation of the hadronic
matrix element we adopt $M=200\,{\rm MeV}$  and $f=92.4\, {\rm MeV}$,
as also used in Ref.~\cite{Fabbrichesi:2017vma}.

Since the underlying flavor-changing magnetic-dipole interaction in
\eq{mdipole} is manifestly $U(1)_D$ gauge invariant, the hadronic
amplitude generated by this interaction inherits the same
gauge-invariant structure also for a massive dark photon. Then, the
amplitude in \eq{M1} can be expressed, as in the massless case, in terms
of a single gauge-invariant dimensionless operator $\hat{O}$ in momentum
space as
\bea
\hat{M}&=&F(z_1.z_2)\,\hat{O}\, ,
\label{CM}
\\
\hat{O}&\equiv&
\frac{1}{m_K^3}\epsilon_{\mu\nu\rho\sigma}
q_1^{\nu}q_2^{\rho}k^{\sigma}\varepsilon_{\D}^{\mu}(k)\, ,
\label{OM}
\eea
where $\epsilon_{\mu\nu\rho\sigma}$ denotes the antisymmetric Levi-Civita
tensor, with $\epsilon_{0123}=1$,
$\varepsilon_{\D}^{\mu}(k)$ the dark-photon polarization vector.
The matching with the explicit computation at one-loop of this amplitude in the $\chi$QM gives the form factor $F(z_1,z_2)$, whose expression is reported in Appendix A. This result extends the corresponding one in \cite{Fabbrichesi:2017vma} to the dark photon mass dependence. The one-loop evaluation of the amplitude in \eq{CM} has been done with the help of Package X \cite{Patel:2015tea,Patel:2016fam}. We adopt the standard Passarino--Veltman notation, with $C_0$ and
$D_0$ denoting the scalar three- and four-point functions,
respectively, and $D_i$ ($i=1,2,3$) and $D_{00}$ the coefficients of the
rank-one and rank-two four-point tensor integrals, respectively.

Finally, the unpolarized square amplitude $|\bar{M}|^2$, summed over all dark-photon polarizations, is given by
\bea
|\bar{M}|^2&=&|F(z_1,z_2)|^2 f_M(z_1,z_2)\, ,
\label{M2PP}
\eea
where
\bea
f_M(z_1,z_2)&=&
\frac{1}{4}\Big[
  4 \Big( z_1 z_2 \left(1-2 z_1 - 2 z_2 \right)
  -\rp \left(z_1 + z_2\right)^2 \Big)
\nonumber\\
  &-& \rDP \Big(4 z_1^2 + \left(1 - 2 z_2\right)^2 +
4 z_1 \left(3 z_2-1\right)\nonumber\\
&+& 4 \rp\left(2 z_1 + 2 z_2-1\right)\Big)
- 2\rDP^2 ( 2 z_1 + 2 z_2 \nonumber\\
 &+& 2 \rp -1) - \rDP^3
\Big]\, ,
\label{fDM}
\eea
and the corresponding differential decay rate is
\bea
\frac{d^2 \Gamma}{d z_1\, dz_2}&=&\frac{m_K}{(4\pi)^3} |\bar{M}|^2\, ,
\label{dGamma1}
\eea
where $\rp=m^2_{\pi}/\mK^2$ and $\rDP=\mDP^2/\mK^2$, with
$\mK$, and $\mDP$, the $K^+$ meson and dark-photon mass respectively. For
$\mpi\simeq 137\, {\rm MeV}$ we used an average mass between $\pi^+$ and $\pi^0$ pion masses since the same approximation is adopted in the loop computation.
In the massless dark photon limit, \eq{fDM} reproduces the
same $f_M(z_1,z_2)$ function in \cite{Fabbrichesi:2017vma}, with
$r^2_1=r^2_2\equiv\rp$.

The kinematic range of $z_{1,2}$ is given by
\be
\sqrt{\rp\rDP} ~\le~ z_1 \le \frac{1}{2}\Big(1-2\sqrt{\rp}-\rDP\Big)\, , ~~~~
z_2^{-} ~\le~ z_2 \le z_2^{+}\, ,
\ee
where
\bea
z_2^{\pm}&=&
\frac{(\rDP + z_1) (1 - \rDP - 2 \rp - 2 z_1)\pm
\sqrt{\Delta}}{2 (\rDP + \rp + 2 z_1)}
\label{kin1}
\\
\Delta&\equiv&(z_1^2-\rDP \rp) (\rDP^2 - 4 \rp + (1 - 2 z_1)^2 
+    \rDP (4 z_1-2))
\nonumber
\eea

Inserting the square amplitude in \eq{M2PP} into the differential decay rate
given in \eq{dGamma1}, and integrating over the allowed phase space, we
obtain the total $\BR$ by normalizing the decay width to the
total charged-kaon width,
$\Gamma_{K^+}^{\rm tot}=5.317\times10^{-14}\ {\rm MeV}$~\cite{ParticleDataGroup:2026mpi}.

In Table~\ref{tableBR1}, we report the corresponding BRs as a function of the
scale $\Lambdasd$, expressed in TeV, for the massless case and for selected
benchmark values of the dark-photon mass. For reference, the last column
shows the BRs obtained by fixing
$\Lambdasd=\bar{\Lambda}_{sd}=2.54\times10^5~{\rm TeV}$, corresponding
to the \KKbar~mixing constraint for a $U(1)_D$ fine structure constant $\alphaD=0.1$, as discussed in
Sec.~\ref{sec:KKconstraints} in \eq{LambdaKKbound}. Here, this value is used only as a reference
normalization for the effective scale. The impact of the different
phenomenological constraints on $\Lambdasd$, and their dependence on the
dark-photon mass, will be analyzed separately in
Sec.~\ref{sec:KKconstraints}. Since the $\BR$ scales as $\Lambdasd^{-2}$, the results can be readily rescaled to any other value of $\Lambdasd$.

For a massless dark photon, the result can be directly compared with
Ref.~\cite{Fabbrichesi:2017vma}. In that analysis, a multiplicative QCD
renormalization factor $\eta=0.5$ was included in the magnetic-dipole
operator. Here we adopt the more conservative choice $\eta=1$, since the
complete set of radiative corrections relevant for the effective operator
is not known. We therefore refrain from including a partial QCD
renormalization effect whose numerical impact could be modified by the
missing higher-order contributions.

\begin{table}[t]
\centering
\begin{tabular}{cc@{\hspace{1.2cm}}c}
\hline\hline
\rule{0pt}{3.0ex}
$\mDP~[{\rm MeV}]$ & ~~$Y~[{\rm TeV}^{-2}]$ &
$\BR\;(\bar{\Lambda}_{sd})$ \\
\hline
\rule{0pt}{2.4ex}
$0$   & $4.61\times10^{4}$ & $7.14\times10^{-7}$ \\
$5$   & $4.52\times10^{4}$ & $7.01\times10^{-7}$ \\
$10$  & $4.40\times10^{4}$ & $6.82\times10^{-7}$ \\
$30$  & $3.69\times10^{4}$ & $5.72\times10^{-7}$ \\
$50$  & $2.83\times10^{4}$ & $4.39\times10^{-7}$ \\
$80$  & $1.66\times10^{4}$ & $2.57\times10^{-7}$ \\
$100$ & $9.75\times10^{3}$ & $1.51\times10^{-7}$ \\
$150$ & $1.37\times10^{3}$ & $2.12\times10^{-8}$ \\
$200$ & $1.07\times 10^{1}$ & $1.56\times10^{-10}$ \\
\hline\hline
\end{tabular}
\caption{Values of
$\BR(K^+\to\pi^+\pi^0\,\gammaD)=Y/(\Lambdasd[{\rm TeV}])^2$
for different values of the dark-photon mass $\mDP$, where the second
column reports the corresponding coefficient $X$ and
$\Lambdasd[{\rm TeV}]$ denotes the numerical value of $\Lambdasd$
expressed in TeV units. The last column shows the upper bound on the
$\BR$, as required by the \KKbar~mixing constraints,
corresponding to the scale
$\bar{\Lambda}_{sd}=2.54\times10^5~{\rm TeV}$ for $\alphaD=0.1$.}
\label{tableBR1}
\end{table}

\subsection{One-pion $K^+\to \pi^+ \gamma\, \gammaD$ radiative decay }
\label{sec:Kdp2}

The effective Hamiltonian in \eq{Hmdipole} also induces the radiative
decay
\bea
K^+(p)\to \pi^+(q_1)\,\gamma(q_2)\,\gammaD(k)\, ,
\eea
where $p$, $q_1$, $q_2$, and $k$ denote the four-momenta of the
corresponding particles.
We introduce, as in the previous $K^+\to \pi^+ \pi^0 \, \gammaD$ process,
the two dimensionless variable $z_{1,2}$ defined as
\bea
z_{1}=\frac{k\cdot q_1}{m_K^2}\, ~~~~z_{2}=\frac{k\cdot q_2}{m_K^2}\, .
\eea

\begin{figure*}[t]
\centering
\includegraphics[width=1.8\columnwidth]{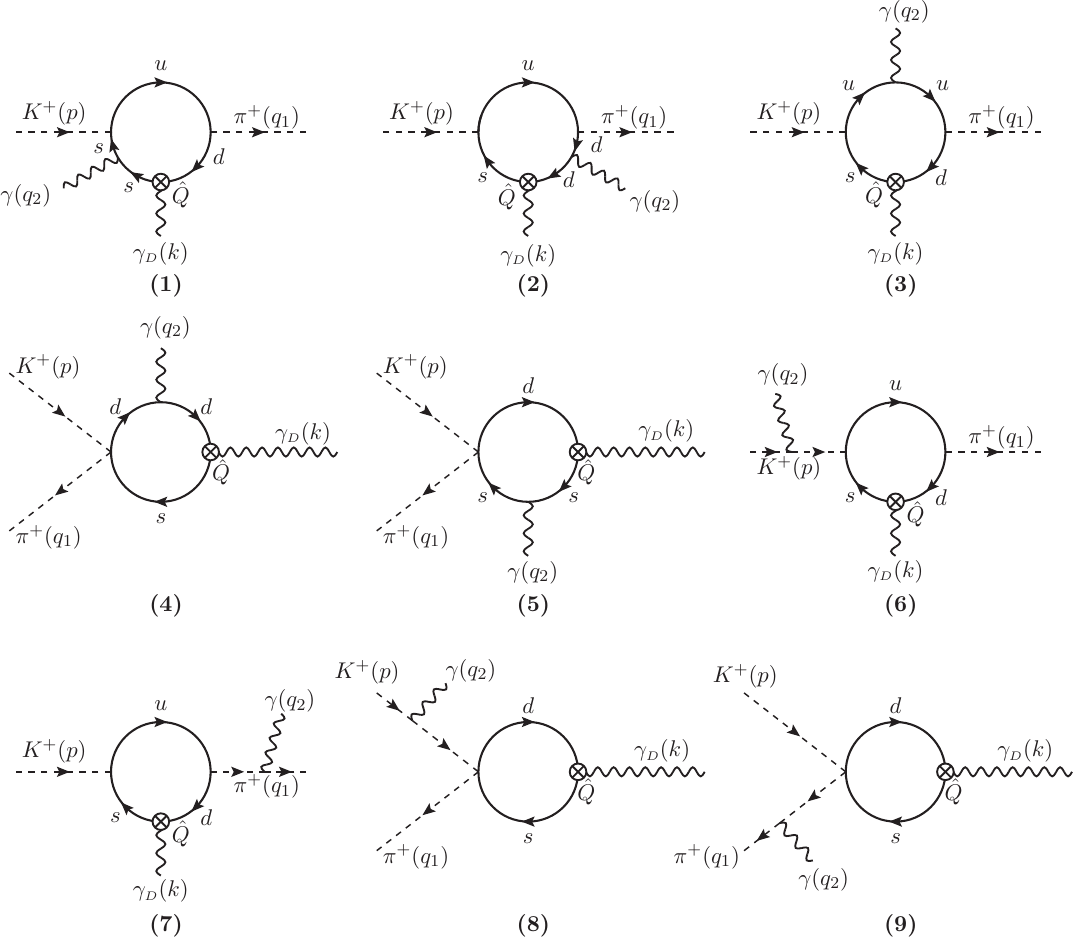}
\caption{\small $\chi$QM diagrams for the process
  $K^+\to\pi^+\,\gamma\,\gammaD$. The crossed circle denotes the insertion
of the FCNC magnetic-dipole operator $\hat Q$ in \eq{mdipole}.}
\label{diagrams2}
\end{figure*}

Following the same approach as for the
$K^+\to\pi^+\pi^0\,\gammaD$ decay, we evaluate the amplitude
\be
\hat{M}\equiv
\langle \pi^+\,\gamma\,\gammaD|
{\cal H}^{\Delta S=1}_{\rm eff}|K^+\rangle
\label{M2}
\ee
by computing the hadronic matrix element
$\langle \gammaD\,\gamma\,\pi^+|\hat{Q}|K^+\rangle$
within the chiral quark model ($\chi$QM)
~\cite{Nishijima:1959,Gursey:1959yy,Gursey:1961,Cronin:1967jq,
Weinberg:1978kz,Manohar:1983md,Manohar:1984uq}.
The one-loop Feynman diagrams contributing to this amplitude are shown
in Fig.~\ref{diagrams2}, namely diagrams (1)-(9).
The evaluation of these diagrams has been done with the help of \textsc{Package X}~\cite{Patel:2015tea,Patel:2016fam}.

Some comments regarding these diagrams are in order.
Contrary to the diagrams contributing to
$K^+\to\pi^+\pi^0\,\gammaD$ in Fig.~\ref{diagrams1}, which are finite,
the diagrams (4) and (5) in Fig.~\ref{diagrams2}, containing the
two-pion--quark vertex, give rise to UV-divergent contributions.
We regularize the one-loop integrals using dimensional regularization
in $d=4-2\varepsilon$ dimensions and adopt the
$\overline{\rm MS}$ subtraction prescription.
The divergent contribution is cancelled by the corresponding counterterm in the higher order chiral Lagrangian.
After the subtraction, the renormalized amplitude retains a logarithmic
dependence on the renormalization scale $\mu$. This is the standard
procedure adopted in chiral perturbation theory for UV-divergent
loop contributions and can be naturally implemented also in the $\chi$QM, which reproduces the same low-energy chiral structure.
We choose the renormalization scale to be of the order of the
chiral-symmetry-breaking scale,
$\mu=\Lambda_{\rm ch}\simeq 1~{\rm GeV}$.

For a generic massive dark photon, the resulting one-loop amplitude can be
matched onto an effective theory involving three independent operators,
which are separately manifestly gauge invariant under QED and
$U(1)_D$.
In particular, the amplitude can be decomposed as
\bea
\hat{M}&=& \FA(z_1,z_2) \,\hat{O}_A+  \FB(z_1,z_2) \, \hat{O}_B+  \FC(z_1.z_2)
\,\hat{O}_C~~~~~~
\label{OM2}
\eea
where the form factors $F_{A,B,C}$, whose expressions are reported in Appendix,
are a function of the Passarino-Veltman functions, and the dimensionless
gauge-invariant operators $\hat{O}_{A,B,C}$ in momentum space are given by
\bea
\hat{O}_A&=&\frac{1}{\mK^2}
\Big[k_{\alpha} q_{2 \mu}-(k\cdot q_2)g_{\mu\alpha}\Big]
\varepsilon_{\gamma}^{\alpha}(q_2)\, \varepsilon_{\D}^{\mu}(k)\, ,
\nonumber\\
\hat{O}_B&=&\frac{1}{m_K^4}\Big[(q_1\cdot q_2) k_{\alpha}q_{1 \mu}-
  (k\cdot q_2) q_{1 \alpha}q_{1 \mu}\nonumber\\
  &+&
  (k\cdot q_1) q_{1 \alpha}q_{2 \mu}
-(k\cdot q_1)(q_1\cdot q_2)g_{\mu\alpha}\Big]
\varepsilon_{\gamma}^{\alpha}(q_2)\, \varepsilon_{\D}^{\mu}(k)\, ,
\nonumber\\
\hat{O}_C&=&\frac{1}{\mK^2}\Big[k_{\alpha} k_{\mu}-\mDP^2 g_{\mu\alpha}
  -\frac{(q_2\cdot k)}{(q_1\cdot q_2)} q_{1\alpha}k_{\mu}
  \nonumber\\
    &+&\frac{\mDP^2}{(q_1\cdot q_2)}q_{1\alpha}q_{2\mu}\Big]
\varepsilon_{\gamma}^{\alpha}(q_2)\, \varepsilon_{\D}^{\mu}(k)\, ,
\label{O2}
\eea
where $\varepsilon_{\D}^{\mu}(k)$ and $\varepsilon_{\gamma}^{\alpha}(q_2)$
are the dark-photon and photon polarization vectors, respectively.
The three operators in \eq{O2} are separately invariant under QED and
$U(1)_D$ gauge transformations, as can be readily verified in momentum
space through the replacements
$\varepsilon_{\gamma}^{\alpha}(q_2)\to
\varepsilon_{\gamma}^{\alpha}(q_2)+q_2^{\alpha}$ and
$\varepsilon_{\D}^{\mu}(k)\to
\varepsilon_{\D}^{\mu}(k)+k^{\mu}$, respectively.
The last operator $\hat{O}_C$
does not contribute in the massless dark photon limit,
as $F_C$ vanishes for $\mDP\to0$.

For an on-shell massive dark photon, the transversality condition
$k_\mu\varepsilon_{\D}^{\mu}(k)=0$ would allow one to eliminate some
terms from the operator basis. Nevertheless, the underlying
flavor-changing magnetic-dipole interaction is manifestly $U(1)_D$
gauge invariant, since it is written in terms of the dark-photon
field-strength tensor. Consequently, the hadronic amplitude generated
by this interaction inherits the same gauge-invariant structure. We
therefore retain the operators in \eq{O2} in a manifestly
gauge-invariant form also for a massive dark photon. In particular,
some terms in $\hat{O}_C$ are redundant on shell because
$k_\mu\varepsilon_{\D}^{\mu}(k)=0$  but are kept to make the
$U(1)_D$ gauge invariance of the operator basis explicit.

The UV-divergent contribution discussed above is entirely
proportional to $\hat{O}_A$ and therefore renormalizes only its
coefficient. Consequently, after the $\overline{\rm MS}$ subtraction,
the residual logarithmic dependence on the renormalization scale $\mu$
is absorbed into the renormalized form factor $F_A(z_1,z_2)$, while the form
factors $F_B(z_1,z_2)$ and $F_C(z_1,z_2)$ are UV finite.

As anticipated above, the contributions from diagrams (6) and (7),
corresponding to initial- and final-state photon radiation, respectively,
vanish in the massless dark photon limit, $\mDP\to0$, as a consequence
of angular-momentum conservation.

We also find that diagrams (8) and (9) in Fig.~\ref{diagrams2},
in which the ordinary photon is emitted from the external charged-meson
lines, give separately vanishing contributions to the amplitude.
Indeed, the corresponding fermion loops contain the antisymmetric
tensor $\sigma^{\mu\nu}$, while only one independent external momentum
flows through the loop. It is therefore not possible to construct a
nonvanishing antisymmetric rank-two tensor, and each of these
contributions vanishes identically.

Finally, by summing over all photon and dark-photon polarizations,
the unpolarized square amplitude in \eq{OM}  is given by
\bea
\bar{M}^2&=&\fAA(z_1,z_2) \,|\FA|^2 +\fBB(z_1,z_2) \, |\FB|^2
\nonumber\\
&+&\fCC(z_1,z_2)|\FC|^2 +
2\Big(\fAB(z_1,z_2)\, {\rm Re}[\FA\FB^*]
\\
&+&\fAC(z_1,z_2)\, {\rm Re}[\FA\FC^*]
+ \fBC(z_1,z_2)\, {\rm Re}[\FB\FC^*]\Big)\, ,
\nonumber
\label{M2gamma}
\eea
where the functions $f_{ij}(z_1,z_2)$, with $\{i,j\}=\{A,B,C\}$, are given by
\bea
\fAA(z_1,z_2)&=&2z_2^2\, ,
\nonumber\\
\fBB(z_1,z_2)&=&
\rp z_2 \Big(\rp z_2 + R z_1\Big)
+\frac{R^2}{4} \Big(\rDP \rp + 2 z_1^2\Big)\, ,
\nonumber\\
\fCC(z_1,z_2)&=& 
\rDP^2 - \frac{4\rDP z_2}{R^2}
\Big(\rp z_2 + R z_1\Big)
\nonumber\\
\fAB(z_1,z_2)&=& \frac{1}{4} \Big(\rDP^3 + 4 \rp z_2^2 + \rDP (R^2-\rDP^2)\Big)\, ,
\nonumber\\
\fAC(z_1,z_2)&=&2 \rDP z_2
\nonumber\\
\fBC(z_1,z_2)&=&-R \rDP z_1\, ,
\label{fABC}
\eea
where $R\equiv  \rDP + \rp + 2 z_1 + 2 z_2-1$, and $\rp=m^2_{\scriptscriptstyle{\pi^+}}/m^2_{\scriptscriptstyle{K}}$. For notational simplicity, the $(z_1,z_2)$ dependence of the form factors $F_{A,B,C}$ is left implicit throughout the paper.

The  corresponding differential decay rate is
\bea
\frac{d^2 \Gamma}{d z_1\, dz_2}&=&\frac{m_K}{(4\pi)^3} |\bar{M}|^2\, ,
\label{dGamma2}
\eea
with $|\bar{M}|^2$ given in \eq{M2gamma}, where
$\rp=m^2_{\pi^+}/\mK^2$ and $\rDP=\mDP^2/\mK^2$, where $m_{\pi^+}$ is
the charged pion mass.

The kinematic range of $z_{1,2}$ is given by
\be
\sqrt{\rp\rDP} ~\le~ z_1 \le \frac{1}{2}\Big(1-\rp-\rDP\Big)\, , ~~~~
z_2^{-} ~\le~ z_2 \le z_2^{+}\, ,
\ee
where
\bea
z_2^{\pm}&=&
\frac{(1 - \rDP - \rp -2 z_1)\left(\rDP+z_1 \pm
\sqrt{z_1^2-\rp \rDP}\right)}{2 (\rDP +\rp +2z_1)}\, . ~~~~~
\label{kin2}
\eea

We now discuss the infrared (IR) behavior of the decay width. This issue
arises only for a massive dark photon, for which a finite photon-energy
cut is required in order to define a finite and experimentally meaningful
exclusive $\BR$.

An important feature of the decay
$K^+\to\pi^+\gamma\,\gammaD$ for a massive dark photon is the presence
of an IR singularity in the soft-photon region. This singularity
is associated with the emission of the ordinary photon from the charged
external states and is therefore of the standard bremsstrahlung type.
It should not be confused with a collinear singularity, since the
nonvanishing kaon and pion masses regulate the collinear region.

In the limit in which the photon momentum $q_2$ becomes soft, the
amplitude factorizes according to the universal soft-photon theorem~\cite{Low:1958sn,Weinberg:1965nx},
\bea
{\cal M}(K^+\to\pi^+\gamma\,\gammaD)\!\!\!&&\underset{E_\gamma\to0}{\simeq}
e\,{\cal M}_0(K^+\to\pi^+\gammaD)~~~~~~~~~~
\nonumber
\\
&&\times
\left[
\frac{q_1\cdot\varepsilon_\gamma}{q_1\cdot q_2}
-\frac{p\cdot\varepsilon_\gamma}{p\cdot q_2}
\right]
+{\cal O}(E_\gamma^0)\, ,
\label{Ampfact}
\eea
where ${\cal M}_0(K^+\to\pi^+\gammaD)$ denotes the amplitude for the
corresponding non-radiative two-body decay, and $e$ the unit of electric charge.
Consequently, the differential decay rate behaves as
\bea
d\Gamma(K^+\to\pi^+\gamma\,\gammaD)
\underset{E_\gamma\to0}{\sim}
\frac{dE_\gamma}{E_\gamma} ,
\eea
and the decay width integrated down to vanishing photon energy is
logarithmically divergent.

In terms of the dimensionless variables $z_1$ and $z_2$, the photon
energy in the kaon rest frame is
\bea
E_\gamma
&=&
\frac{\mK^2-\mpi^2-\mDP^2-2\mK^2z_1}{2\mK}
\nonumber\\
&=&
\mK\left(z_1^{\rm max}-z_1\right).
\eea
The soft-photon region therefore corresponds to
$z_1\to z_1^{\rm max}$, together with $z_2\to0$. Accordingly, close to
the endpoint the differential rate develops the characteristic behavior
\bea
\frac{d\Gamma}{dz_1}
\underset{z_1\to z_1^{\rm max}}{\sim}
\frac{1}{z_1^{\rm max}-z_1},
\eea
which generates a logarithmic divergence upon integration.

A finite and experimentally meaningful $\BR$ must therefore
be defined by imposing a lower cut on the photon energy,
\bea
E_\gamma>E_\gamma^{\rm cut}.
\eea
In terms of $z_1$, this condition is equivalent to
\bea
z_1<
z_1^{\rm max}-\frac{E_\gamma^{\rm cut}}{\mK}.
\label{z1max}
\eea
The leading dependence of the integrated rate on the cut is logarithmic,
\bea
\Gamma(E_\gamma>E_\gamma^{\rm cut})
=
A(\mDP)\,
\log\!\left(\frac{E_0}{E_\gamma^{\rm cut}}\right)
+\Gamma_{\rm finite}
+{\cal O}(E_\gamma^{\rm cut}),
\nonumber
\eea
where $E_0$ is a scale of the order of the characteristic energy
available in the decay, while the coefficient $A(\mDP)$ is fixed by the
soft limit of the amplitude in \eq{Ampfact}

Notice that this IR behavior is specific to a massive dark photon.
Indeed, the coefficient of the soft singularity is proportional to the
non-radiative decay rate $\Gamma(K^+\to\pi^+\gammaD)$. For the magnetic-dipole
interaction considered here, the corresponding two-body amplitude
vanishes in the massless dark photon limit. Hence the coefficient of the
soft logarithm vanishes for $\mDP\to0$, consistently recovering a finite
$K^+\to\pi^+\gamma\,\gammaD$ rate in that limit.

To estimate a realistic value of the photon-energy cut, we consider the
NA62 experimental conditions \cite{NA62:2017rwk}. The charged-kaon beam has an energy of
about $75~{\rm GeV}$, corresponding to a Lorentz factor
$\gamma_K\simeq 152$, while the large-angle photon-veto system is
designed to detect photons with laboratory energies down to about
$200~{\rm MeV}$.
For a photon emitted at an angle $\theta_\gamma$ with respect to the
kaon boost direction, as measured in the kaon rest frame, the laboratory
and kaon-rest-frame photon energies are related by
\bea
E_\gamma^{\rm lab}
=
\gamma_K E_\gamma
\left(1+\beta_K\cos\theta_\gamma\right),
\eea
where $\beta_K$ and $\gamma_K$ are the velocity and Lorentz factor of
the kaon in the laboratory frame, respectively.
For photons emitted approximately along the kaon flight direction,
$\cos\theta_\gamma\simeq 1$, so that
\bea
E_\gamma^{\rm lab}
\simeq
\gamma_K E_\gamma(1+\beta_K)
\simeq
2\gamma_K E_\gamma .
\nonumber
\eea
Using the NA62 value $\gamma_K\simeq 152$, a photon with
$E_\gamma^{\rm lab}\simeq 200~{\rm MeV}$ corresponds, in the kaon rest
frame, to
\bea
E_\gamma
\simeq
\frac{E_\gamma^{\rm lab}}{2\gamma_K}
\simeq
0.66~{\rm MeV}.
\eea
This indicates that a cut of order
$E_\gamma^{\rm cut}\sim 0.5$--$0.7~{\rm MeV}$ in the kaon rest frame
provides a realistic benchmark for the experimental photon-energy
sensitivity~\cite{NA62:2017rwk}. We therefore adopt
$E_\gamma^{\rm cut}=0.5~{\rm MeV}$ as a representative benchmark.
Using $E_\gamma=\mK(z_1^{\rm max}-z_1)$, this corresponds to
\bea
z_1^{\rm max}-z_1
\simeq
\frac{0.5~{\rm MeV}}{\mK}
\simeq 10^{-3},
\eea
namely a cut at the per-mille level near the kinematic endpoint.

For definiteness, in the numerical analysis we implement the photon-energy
cut through a common prescription for both the massless and massive
dark photon cases. Namely, we replace the kinematic endpoint
$z_1^{\rm max}$ by the cut endpoint
\bea
z_{1,{\rm cut}}^{\rm max}
=
z_1^{\rm max}(1-\delta),
\qquad
\delta=10^{-3}\, .
\label{z1cut}
\eea
We use this prescription throughout the numerical evaluation of the $\BR$s.

Inserting the amplitude in \eq{M2gamma} into the differential decay rate
given in \eq{dGamma2}, and integrating over the allowed phase space with
a cut in the photon energy as in \eq{z1cut}, we
obtain the total $\BR$ by normalizing the decay width to the
total charged-kaon width.

In Table~\ref{tableBR2}, we report the corresponding BRs as a function of the
scale $\Lambdasd$, expressed in TeV, for the massless case and for selected
benchmark values of the dark-photon mass. As in Table~\ref{tableBR1}, in the last column, we show the upper bounds on the BRs allowed by the \KKbar~mixing.

\begin{table}[t]
\centering
\begin{tabular}{cc@{\hspace{1.2cm}}c}
\hline\hline
\rule{0pt}{3.0ex}
$\mDP~[{\rm MeV}]$ & ~~$Y~[{\rm TeV}^{-2}]$ &
$\BR\;(\bar{\Lambda}_{sd})$ 
\\[0.7ex]
\hline
\rule{0pt}{2.9ex}
$0$   & $4.08\times10^{1}$ & $6.32\times10^{-10}$ \\
$5$   & $4.49\times10^{1}$ & $6.96\times10^{-10}$ \\
$10$  & $5.64\times10^{1}$ & $8.75\times10^{-10}$ \\
$30$  & $1.77\times10^{2}$ & $2.74\times10^{-9}$ \\
$50$  & $4.09\times10^{2}$ & $6.34\times10^{-9}$ \\
$80$  & $9.20\times10^{2}$ & $1.43\times10^{-8}$ \\
$100$ & $1.33\times10^{3}$ & $2.06\times10^{-8}$ \\
$150$ & $2.29\times10^{3}$ & $3.55\times10^{-8}$ \\
$200$ & $2.66\times10^{3}$ & $4.13\times10^{-8}$ \\
\hline\hline
\end{tabular}
\caption{Values of
$\BR(K^+\to\pi^+\gamma\,\gammaD)=Y/(\Lambdasd[{\rm TeV}])^2$
for different values of the dark-photon mass $\mDP$, where the second
column reports the corresponding coefficient $X$ and
$\Lambdasd[{\rm TeV}]$ denotes the numerical value of $\Lambdasd$
expressed in TeV units. The last column shows the upper bound on the $\BR$,
as required by the \KKbar~mixing constraints,
corresponding to the scale
$\bar{\Lambda}_{sd}=2.54\times10^5~{\rm TeV}$ for $\alphaD=0.1$.}
\label{tableBR2}
\end{table}

\subsection{One-pion $K^+\to \pi^+ \gammaD$ non-radiative decay }
\label{sec:Kdp3}

We now analyze the decay $K^+\to\pi^+\gammaD$, which is relevant
only for a massive dark photon, since the corresponding two-body
decay into a massless dark photon is forbidden by angular-momentum
conservation. This process gives rise to the signature of a resonant
$K^+\to\pi^+X$ decay, with $X$ appearing as missing energy and
producing a peak in the missing-mass distribution at $m_X$.
This process has recently been searched for by the
NA62 experiment~\cite{NA62:2025upx}. In the absence of a signal, strong
constraints on the $\BR$ of this process have been set in the mass range $1~{\rm MeV}\le m_X \le 261~{\rm MeV}$. We will therefore use the
$K^+\to \pi^+ \gammaD$ decay to derive constraints on the scale
$\Lambdasd$ for dark-photon masses in the same range. The corresponding
results will be presented in the next section.

The Hamiltonian in \eq{Hmdipole} can also directly induce the decay
\bea
K^+(p)\to \pi^+(q)\, \gammaD(k)\, .
\eea
In this case, the amplitude in momentum space is given by
\bea
M&=&\frac{1}{\Lambdasd}\, X_{\mu\nu}(k^2)\,
\varepsilon^{\mu}_{\scriptscriptstyle{D}}(k) k^{\nu}\, ,
\eea
where
$X_{\mu\nu}(k^2)\equiv
\langle \pi^+(q)|\,(\bar{s}\sigma_{\mu\nu} d)\,|K^+(p)\rangle$
is the hadronic matrix element of the local antisymmetric tensor operator
$(\bar{s}\sigma_{\mu\nu} d)$ evaluated at the origin, which depends on $k^2=\mDP^2$, and $\varepsilon^{\mu}_{\scriptscriptstyle{D}}(k)$ is the dark-photon
polarization vector.

\begin{figure}[t]
\centering
\includegraphics[width=0.55\columnwidth]{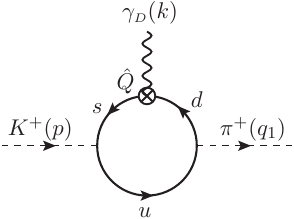}
\caption{\small $\chi$QM diagram for the process
$K^+\rightarrow\pi^+\,\gammaD$. The crossed circle stands for
  the insertion of the FCNC magnetic-dipole operator $\hat Q$ in
\eq{mdipole}.}
\label{diagram3}
\end{figure}

In the $\chi$QM, only one Feynman diagram contributes to this process, as
shown in Fig.~\ref{diagram3}.

The corresponding result for $X_{\mu\nu}$ is
\bea
X_{\mu\nu}(k^2)&=&\Big(q^{\mu}p^{\nu}-q^{\nu}p^{\mu}\Big) F_X(k^2)
\label{XChQM}
\eea
where the scalar form factor is
\bea
F_X(k^2)=\frac{iM^3}{2\pi^2 f^2}C_0\Big(\mDP^2,\mK^2,\mpi^2;M,M,M\Big)\, ,
\eea
where $f=92.4~{\rm MeV}$ is the pion decay constant used in $\chi$QM~\cite{Antonelli:1995nv}.

However, the corresponding tensor matrix element for the FCNC operator
in \eq{mdipole} between $K^0$ and $\pi^0$ has also been computed using
lattice QCD in Ref.~\cite{Baum:2011rm}. Using isospin symmetry to relate
the $K^+$ and $\pi^+$ matrix element to the neutral one, and following
the parametrization adopted in Ref.~\cite{Baum:2011rm}, we obtain
\bea
X_{\mu\nu}(k^2)=\Big(k^{\mu}p^{\nu}-k^{\nu}p^{\mu}\Big)
\frac{2 f_T^{K\pi}(k^2)}{\mK+\mpi}\, ,
\label{XQCD}
\eea
where the tensor form factor $f_T^{K\pi}(k^2)$ given in
Ref.~\cite{Baum:2011rm} is parametrized as
\bea
f_T^{K\pi}(k^2)&=& \frac{f_T^{K\pi}(0)}{1-s_T^{K\pi} k^2}\, ,
\eea
with~\cite{Baum:2011rm}
\bea
f_T^{K\pi}(0)&=&0.417\pm 0.015\, ,
\nonumber\\
s_T^{K\pi}&=&(1.10\pm 0.14)\, {\rm GeV}^{-2}\, .
\eea
By matching the two results in \eq{XChQM} and \eq{XQCD}, up to an
overall phase that does not play any role here, we find
\bea
f_T^{K\pi}(\mDP^2)&=&\frac{\left(\mK+\mpi\right) M^3}{4\pi^2 f^2}
\nonumber\\
&\times&
\Big|C_0\Big(\mDP^2,\mK^2,\mpi^2;M,M,M\Big)\Big|\, ,
\label{FTrel}
\eea
where the Passarino--Veltman notation for the $C_0$ functions is that introduced in Sec.\ref{sec:Kdp1}. 
The relation in \eq{FTrel}  provides a direct connection between the $\chi$QM prediction and the lattice-QCD result for the hadronic matrix element of $\hat{Q}$.
We find that this relation is satisfied within a few percent over the entire
kinematically allowed range of $\mDP$, for the value
$M=200~{\rm MeV}$ adopted in our analysis. This agreement is well
within the $2\sigma$ uncertainty on $f_T^{K\pi}(0)$.

The good agreement in \eq{FTrel} strengthens our confidence in the
validity of the $\chi$QM approach adopted here to evaluate the hadronic
matrix elements of the other processes, for which the effective
constituent-quark mass $M=200~{\rm MeV}$ is used. Nevertheless, whenever
available, we prefer to rely on first-principles lattice-QCD calculations
of the relevant hadronic matrix elements. Therefore, in the numerical
analysis of the $K^+\to\pi^+\gammaD$ decay, we use the central values of
the lattice-QCD determination of the corresponding tensor form factor
given in Ref.~\cite{Baum:2011rm}.

By summing over all dark-photon polarizations, the squared amplitude is
\bea
|\bar{M}|^2&=& \frac{4 \mK^4 \mDP^2}{\Lambdasd^2}
\Big[ \frac{1}{4}\left(1-\rp-\rDP\right)^2-\rp\rDP\Big]
\nonumber\\
&\times&
\left(\frac{f_T^{K\pi}(k^2)}{\mK+\mpi}\right)^2\, .
\label{M2piDP}
\eea
As expected, the squared amplitude in \eq{M2piDP} vanishes in the
massless dark photon limit, $m_{\gammaD}\to 0$, consistently with
angular-momentum conservation.
The total decay width is then
\bea
\Gamma &=& \frac{1}{16\pi \mK^3}|\bar{M}|^2
\sqrt{\lambda(\mK^2,\mpi^2,\mDP^2)}\, ,
\eea
where $\lambda(a,b,c)$ denotes the K\"all\'en function.

\subsection{Kinetic-mixing contribution}
\label{sec:kinetic-mixing}
  
Here we consider the two-pion $K^+\to\pi^+\pi^0\,\gammaD$ and the
radiative one-pion $K^+\to\pi^+\gamma\,\gammaD$ decays induced purely
by photon--dark-photon kinetic mixing in the massive dark-photon
scenario.

A kinetic-mixing term between the photon and the dark photon is
typically parametrized in the Lagrangian as \cite{Holdom:1985ag,Fabbrichesi:2020wbt}
\bea
{\cal L}_{\rm kin-mixing}
&=&
-\frac{\varepsilon}{2} F_{\mu\nu}F_{\scriptscriptstyle{D}}^{\mu\nu}\,,
\label{kmixing}
\eea
where $F_{\mu\nu}$ denotes the photon field strength.
Even if absent at tree level, the term in \eq{kmixing} can be induced
by radiative corrections in the presence of messenger fields or other
portal states that communicate between the dark and visible SM sectors.

In the massless dark photon case, the kinetic-mixing term can be
removed by a field redefinition and absorbed into millicharge
interactions of dark-sector matter with SM charged fields, leaving no
direct renormalizable coupling of the dark photon to SM fields.
In contrast, for a massive dark photon the kinetic mixing cannot be
eliminated from the physical interactions. After diagonalizing the
kinetic terms, electrically charged SM fields acquire a coupling to
the dark photon proportional to $e\varepsilon$, described by the Lagrangian
\cite{Fabbrichesi:2020wbt}
\bea
 {\cal L} &\supset& -\frac{\sqrt{e\varepsilon}}{\sqrt{1-\varepsilon^2}}
J_{\mu} A^{\mu}_{\scriptscriptstyle{D}}\simeq -e\, \varepsilon \,
J_{\mu} A^{\mu}_{\scriptscriptstyle{D}}\, ,
\label{LDPint}
\eea
where $J_{\mu}$ is electromagnetic current of the ordinary SM matter, and $e$ denotes the unit of electric charge. This is the usual dimension-4 interaction of the massive dark photon via the QED current.

Therefore, for a sufficiently small but nonvanishing dark-photon mass,
by rescaling the usual photon coupling as $e\to\epsilon e$ in the SM
amplitudes for the decays $K^+\to\pi^+\pi^0\gamma$ and
$K^+\to\pi^+\gamma\gamma$, one can estimate the corresponding BRs for
$K^+\to\pi^+\pi^0\,\gammaD$ and $K^+\to\pi^+\gamma\,\gammaD$,
respectively, as
\bea
{\rm BR}(K^+\to\pi^+\pi^0\,\gammaD)&\simeq&\varepsilon^2{\rm BR}(K^+\to\pi^+\pi^0\,\gamma)~~~~~
\nonumber\\
    {\rm BR}(K^+\to\pi^+\gamma\,\gammaD)&\simeq&2\varepsilon^2{\rm BR}(K^+\to\pi^+\gamma\,\gamma)\, ,
    \label{BRmixing}
\eea
where the $\BR$s in the right hand side of \eq{BRmixing} correspond to the SM ones.

For larger values of
$\mDP$, however, this approximation progressively loses its validity: a consistent evaluation would require recalculating the corresponding SM amplitudes and phase-space integrals with one of the photons replaced by a massive vector boson,  including
the massive dark photon kinematics, the corresponding modification of
the phase space, and the appropriate experimental acceptance. Such an
analysis is beyond the scope of the present work. We do not perform such a dedicated calculation here, since our purpose is only to estimate the
order of magnitude of the kinetic-mixing contribution.
In particular, the reduction of the available phase space associated
with a finite dark-photon mass is expected to suppress the rates
relative to the massless rescaling, so that the estimates obtained from \eq{BRmixing} for larger values of $\mDP$
are expected to be conservative as far as phase-space effects
are concerned. Possible effects associated with the longitudinal dark-photon
polarization are not captured by this simple rescaling and would
require the full massive-vector calculation.
This level of accuracy is sufficient for our purposes,
since, as shown below, the resulting kinetic-mixing contributions are
already far below the present experimental sensitivity.

The current measurements for the $\BR$s of SM processes 
$K^+\to\pi^+\pi^0\,\gamma$ and $K^+\to\pi^+\gamma\,\gamma$ are~\cite{ParticleDataGroup:2026mpi}
\bea
{\rm BR}(K^+\to\pi^+\pi^0\,\gamma)&=&(6.0\pm 0.4)\times 10^{-6}
\nonumber\\
    {\rm BR}(K^+\to\pi^+\gamma\,\gamma)&=&(1.01\pm 0.06)\times 10^{-6}
    \label{BRexp}
    \eea

Strong constraints on the kinetic-mixing parameter $\varepsilon$ for
a massive dark photon decaying into invisible final states have been
obtained from several experiments, including E787~\cite{E787:2001urh}, E949~\cite{BNL-E949:2009dza}, NA62~\cite{NA62:2019meo},
BaBar~\cite{BaBar:2017tiz}, and NA64 \cite{Banerjee:2019pds,NA64:2023wbi},
 as well as from the anomalous magnetic moments of the electron $(g-2)_e$
and muon $(g-2)_\mu$  \cite{Abdullahi:2023tyk,Serao:2026qar}.
In the dark-photon mass range relevant to our analysis,
the most stringent constraints arise mainly from the updated analysis of
NA64~\cite{NA64:2023wbi}, and BaBar~\cite{BaBar:2017tiz}  bounds.
For an overview of the various constraints in the whole kinematic range of massive dark photon scenario, see, for instance,
Ref.~\cite{Fabbrichesi:2020wbt}.

In Table~\ref{tab:mixing}, we report the upper bounds on $\varepsilon$,
together with the corresponding upper bounds on the $\BR$s
induced by kinetic mixing, for representative benchmark values of the
dark-photon mass. The bounds on $\varepsilon$ in this mass range are
predominantly determined by the updated NA64 results~\cite{NA64:2023wbi}.

For dark-photon masses below approximately $30~{\rm MeV}$, the updated
NA64 constraints require $\varepsilon$ to be of order $10^{-5}$ or
smaller. As a consequence, the corresponding upper bounds on the
$\BR$s in \eq{BRmixing} are at the level of $10^{-15}$ or
below. At larger dark-photon masses, the upper bounds on the $\BR$s increase, reaching values of a few $10^{-13}$ for the
$K^+\to\pi^+\pi^0\gammaD$ channel in the region above
$100~{\rm MeV}$, while remaining at the level of $10^{-14}$ for the
$K^+\to\pi^+\gamma\gammaD$ channel.

For the massive-dark-photon case, where production can also proceed
through kinetic mixing (Sec.~II.D), SN1987A constraints on the mixing
parameter $\varepsilon$ have likewise been derived in the
literature~\cite{Chang:2016ntp}, excluding a band of roughly
$\varepsilon \sim 10^{-10}$--$10^{-6}$ for dark-photon masses up to
$\mathcal{O}(100)$~MeV. These limits can be used to further restrict those in Tab.~III.

Then we conclude that massive dark-photon production through kinetic
mixing in these kaon decay channels is strongly suppressed and would
require sensitivities well beyond those currently achieved at kaon
experiments.

\begin{table}[h!]
\centering
\begin{tabular}{cccc}

\hline\hline
\rule{0pt}{3.9ex}
\!\!\!\!\!
$\mDP[{\rm MeV}]$ &
$\varepsilon_{\rm max}$ &
\!\!\!${\rm BR}\bigl(\text{\small $K^+\!\!\!\to\!\pi^+\pi^0\gammaD$}\bigr)$ &
${\rm BR}\bigl(\text{\small $K^+\!\!\!\to\!\pi^+\gamma\gammaD$}\bigr)$
\\[0.7ex]
\hline
\\[-2ex]
$5$   & $8.0\times10^{-6}$ & $3.8\times10^{-16}$ & $1.3\times10^{-16}$ \\
$10$  & $1.4\times10^{-5}$ & $1.2\times10^{-15}$ & $4.0\times10^{-16}$ \\
$30$  & $3.0\times10^{-5}$ & $5.4\times10^{-15}$ & $1.8\times10^{-15}$ \\
$50$  & $4.8\times10^{-5}$ & $1.4\times10^{-14}$ & $4.7\times10^{-15}$ \\
$80$  & $8.0\times10^{-5}$ & $3.8\times10^{-14}$ & $1.3\times10^{-14}$ \\
$100$ & $1.2\times10^{-4}$ & $8.6\times10^{-14}$ & $2.9\times10^{-14}$ \\
$150$ & $2.0\times10^{-4}$ & $2.4\times10^{-13}$ & $8.1\times10^{-14}$ \\
$200$ & $7.0\times10^{-5}$ & $2.9\times10^{-14}$ & $9.9\times10^{-15}$ \\
\hline\hline
\end{tabular}
\caption{Upper bounds on $\varepsilon$ at $90\%$ C.L.
  from NA64~\cite{NA64:2023wbi}
and NA62~\cite{NA62:2025upx}, and corresponding upper bounds on the
$\BR$s $\BR(K^+\to\pi^+\pi^0\gammaD)$ and
$\BR(K^+\to\pi^+\gamma\gammaD)$ for dark-photon production via
kinetic mixing $\varepsilon$, as in \eq{BRmixing},
for some representative values of dark photon
masses $\mDP$. For $\mDP\gtrsim 10~{\rm MeV}$, the BR bounds should be regarded as order-of-magnitude estimates.}
\label{tab:mixing}
\end{table}

\section{Constraints from  \KKbar~mixing}
\label{sec:KKconstraints}
Here we consider the constraints on the scale $\Lambdasd$, appearing in
the magnetic-dipole interactions in \eq{Hmdipole}, induced by
the \KKbar~mixing. 
To that end, a UV-complete dark-sector model responsible for generating
the magnetic-dipole interaction must be specified.
A minimal realization is obtained by introducing a heavy dark fermion $Q$, singlet under the SM gauge group and charged under $U(1)_D$, together with
scalar messenger fields $S_L$ and $S_R$ carrying both SM and dark quantum
numbers. The interactions of these fields with the SM quarks are given by
\be
{\cal L}_{\rm int} =
g_L(\bar Q_L q_R)S_R+
g_R(\bar Q_R q_L)S_L+{\rm h.c.}\, .
\ee
Here $q_L$ and $q_R$ denote the SM quark fields, transforming as color
triplets and as $SU(2)_L$ doublets and singlets, respectively. Flavor indices
are left implicit, and the couplings $g_L$ and $g_R$ are taken to be
flavor universal. The scalar messenger $S_L$ transforms as an $SU(2)_L$
doublet, whereas $S_R$ is an $SU(2)_L$ singlet; both are color triplets.
In addition, the messenger fields carry $U(1)_D$ charge, chosen to be the
same as that of the dark fermion $Q$.
A mixing between $S_L$ and $S_R$, induced after electroweak and dark-sector
symmetry breaking, allows both quark chiralities to participate in the loop.
The Wilson coefficient multiplying the magnetic dipole operator $\hat{Q}$ in
\eq{Hmdipole} is obtained by integrating the vertex function
in our simplified model (see Fig. \ref{vertex}).

Assuming a common mass scale $\Lambda$ for the dark fermion and the lightest
scalar messenger, corresponding to the maximal chiral enhancement, the
one-loop matching gives
\bea
\frac{1}{\Lambdasd}
&=&
\frac{e_D}{32\pi^2}\,
\frac{\xi}{\Lambda}\, ,
\label{Lambdaxi}
\eea
where $\xi=\frac{g_Lg_R}{2}$ and $e_D$ is the corresponding $U(1)_D$ coupling of dark photon with messengers and dark fermion fields in the dark sector.
We have checked \eq{Lambdaxi} by means of Package X
\cite{Patel:2015tea,Patel:2016fam}.

This relation allows us to express the magnetic-dipole interaction in terms of the parameters of the simplified
dark-sector model and to compare its contribution with the corresponding
new-physics contribution to \KKbar~mixing. Such a comparison is
therefore necessarily model dependent. 
The same scalar--fermion interactions responsible for the
$\Delta S=1$ magnetic-dipole operator also induce
$\Delta S=2$ transitions contributing to \KKbar~mixing.
The corresponding contribution to the neutral-kaon mass difference is
\bea
\Delta M_K
=
2\,{\rm Re}\,
\langle K^0|
{\cal H}_{\rm eff}^{\Delta S=2}
|\bar K^0\rangle .
\label{eq:DMK}
\eea
The effective Hamiltonian contains the complete set of
$\Delta S=2$ four-quark operators,
\bea
{\cal H}_{\rm eff}^{\Delta S=2}
=
\sum_{i=1}^{5} C_i Q_i
+
\sum_{i=1}^{3}\widetilde C_i\widetilde Q_i .
\label{eq:HeffDS2}
\eea
For the basis of operators appearing in \eq{eq:HeffDS2} see
Refs. \cite{Beall:1981ze,Gerard:1984bg,Hagelin:1992tc,Gabbiani:1996hi,Ciuchini:1998ix}. The corresponding Wilson coefficients were computed in
Ref.~\cite{Fabbrichesi:2017vma}, including the ${\cal O}(\alpha_s)$
next-to-leading-order QCD evolution from the matching scale
$\Lambda$ down to $\mu\simeq2~{\rm GeV}$, where the hadronic
matrix elements are evaluated using lattice-QCD results \cite{Ciuchini:1998ix}. 
In the large-mixing limit of the messenger sector, and assuming
equal masses $\Lambda$ for the lightest messenger and dark-fermion
states, one obtains \cite{Fabbrichesi:2017vma}
\bea
\Delta M_K &=& 8.47\times10^{-13}\,
\frac{\xi^2}{\Lambda^2}\, {\rm TeV}
\nonumber\\
&=&6.72\times 10^{-9}\frac{{\rm TeV}}{\alphaD \Lambdasd^2}\, ,
\label{eq:DMKK}
\eea
where $\Lambda$ and $\Lambdasd$ are expressed in TeV, and in the last relation
of \eq{eq:DMKK} we have used \eq{Lambdaxi}, with $\alphaD$ the fine structure
constant of $U(1)_D$.

Following the standard procedure commonly adopted in deriving
constraints from \KKbar~mixing, we allow for the theoretical
uncertainties affecting the SM prediction of $\Delta M_K$,
in particular those associated with nonperturbative long-distance
contributions. We therefore conservatively require the new-physics
contribution not to exceed $30\%$ of the measured central value of
$\Delta M_K$~\cite{ParticleDataGroup:2026mpi}.
This provides a constraint on the effective
scale $\Lambdasd$ as a function of $\alphaD$, in particular
\be
\Lambdasd \ge \bar{\Lambda}_{sd} \, \sim \, 2.54 \times 10^{5}\, {\rm TeV}
\sqrt{\frac{0.1}{\alpha}}\, .
\label{LambdaKKbound}
\ee

The impact of these constraints on the upper bounds of
${\rm BR}(K^+\to\pi^+\pi^0\gammaD)$ and
${\rm BR}(K^+\to\pi^+\gamma\gammaD)$, for $\alphaD=0.1$, is shown
in Tables~\ref{tableBR1} and~\ref{tableBR2} for representative benchmark
values of the dark-photon mass. While in the massive dark-photon
scenario the strongest constraints on the scale $\Lambdasd$ arise from
the experimental bounds on ${\rm BR}(K^+\to\pi^+X)$, in the massless
dark photon case the most stringent constraint is instead provided by
\KKbar~mixing. In particular, the latter implies the upper
bounds
\begin{equation}
\boxed{
\begin{aligned}
{\rm BR}(K^+\to\pi^+\pi^0\gammaD)
&\lesssim 6.98\times10^{-7}\,,\\
{\rm BR}(K^+\to\pi^+\gamma\gammaD)
&\lesssim 6.18\times10^{-10}\,.
\end{aligned}
}
\label{BRmassless}
\end{equation}
which identify the sensitivity range relevant for searches for a
massless dark photon at NA62 and future kaon facilities.

\section{Astrophysical constraints and their limitations}
\label{sec:SN-caveats}

A stronger constraint on the massless dark-photon scenario than the one
derived above from \KKbar~mixing can in principle be inferred
from the SN1987A neutrino signal~\cite{Camalich:2020wac}, which bounds the decay $\Lambda \to n \gammaD$ through the
requirement that dark-photon emission not exceed the observed
neutrino-burst duration. Translated into the present normalization, this
bound corresponds to
\bea
  \mathrm{BR}(K^+ \to \pi^+\pi^0\gammaD)\; &\lesssim& \; 1.28\times10^{-9}\, ,
  \nonumber\\
 \mathrm{BR}(K^+\to\pi^+\gamma\,\gammaD)\; &\lesssim& \; 1.13\times10^{-12}\, ,
\label{eq:SN-bound-2pi}
\eea
roughly three orders of magnitude below the \KKbar~mixing bound of
Eq.~(56). We do not adopt Eq.~\eqref{eq:SN-bound-2pi} as the basis for the
sensitivity targets quoted in this work, for two independent reasons.

\paragraph{Astrophysical systematics.}
The bound relies on the density, temperature, and composition profile
assumed for the proto-neutron-star core, which are not directly observed
but taken from numerical simulations, and are based on the assumption that the
produced dark photons free-stream out of the core without significant
reabsorption. Varying these inputs within their current uncertainties has
been found to shift comparable supernova energy-loss bounds by up to an
order of magnitude~\cite{Rrapaj:2015wgs}. More fundamentally, the
robustness of the standard SN1987A cooling argument itself -- independent
of any dark-sector modeling -- has been questioned in~\cite{Bar:2019ifz,Page:2020gsx}, which point
out that the observed neutrino signal remains compatible with cooling
scenarios in which anomalous energy losses play a significantly larger
role than assumed in the fiducial neutrino-driven picture. We therefore
regard Eq.~\eqref{eq:SN-bound-2pi} as indicative rather than as a firm
constraint.

\paragraph{Restriction to the massless case.}
Equation~\eqref{eq:SN-bound-2pi} is derived in~\cite{Camalich:2020wac} specifically for
a massless dark photon, produced thermally in the core via
$\Lambda \to n\gammaD$ and assumed to escape as radiation. No
corresponding calculation exists in the literature for a massive
dark photon in this channel. Extending the bound to $m_{\gammaD}\neq0$ is
not simply a matter of rescaling the two-body decay phase space: the
astrophysical process additionally involves (a) the thermal population of
$\Lambda$ hyperons available to decay, which is Boltzmann-suppressed once
$m_{\gammaD}$ becomes comparable to the core temperature,
$T\sim30$--$50$~MeV, and (b) the possibility that a sufficiently massive,
or sufficiently strongly coupled, dark photon is trapped within the core
rather than free-streaming out of it -- a regime in which the bound would
weaken rather than strengthen. In the absence of a dedicated calculation
of these effects, any use of the SN1987A bound of Ref.~\cite{Camalich:2020wac}
should be restricted to $m_{\gammaD}=0$, where it was actually derived, and not extrapolated into the massive-dark-photon region of parameter space explored in Sec.~IV.

For these reasons, we regard the \KKbar~mixing constraint derived above
as providing the strongest laboratory-based bound relevant for the
massless dark photon sensitivity targets discussed in Sec.~V, while
the SN1987A bound in Eq.~\eqref{eq:SN-bound-2pi} is quoted only for
comparison and is restricted to $m_{\gammaD}=0$.

\section{Phenomenological bounds}
\label{sec:phenobounds}

We now discuss the phenomenological implications of the results derived
in the previous sections for the two main decay channels
$K^+\to\pi^+\pi^0\gammaD$ and $K^+\to\pi^+\gamma\gammaD$ induced by
the FCNC magnetic-dipole operator in \eq{mdipole}. In particular, we
analyze their dependence on the dark-photon mass and the upper bounds
on their $\BR$s obtained by taking into account the
constraints on the effective scale $\Lambdasd$ discussed above.

\begin{figure}[t]
\centering
\includegraphics[width=1\columnwidth]{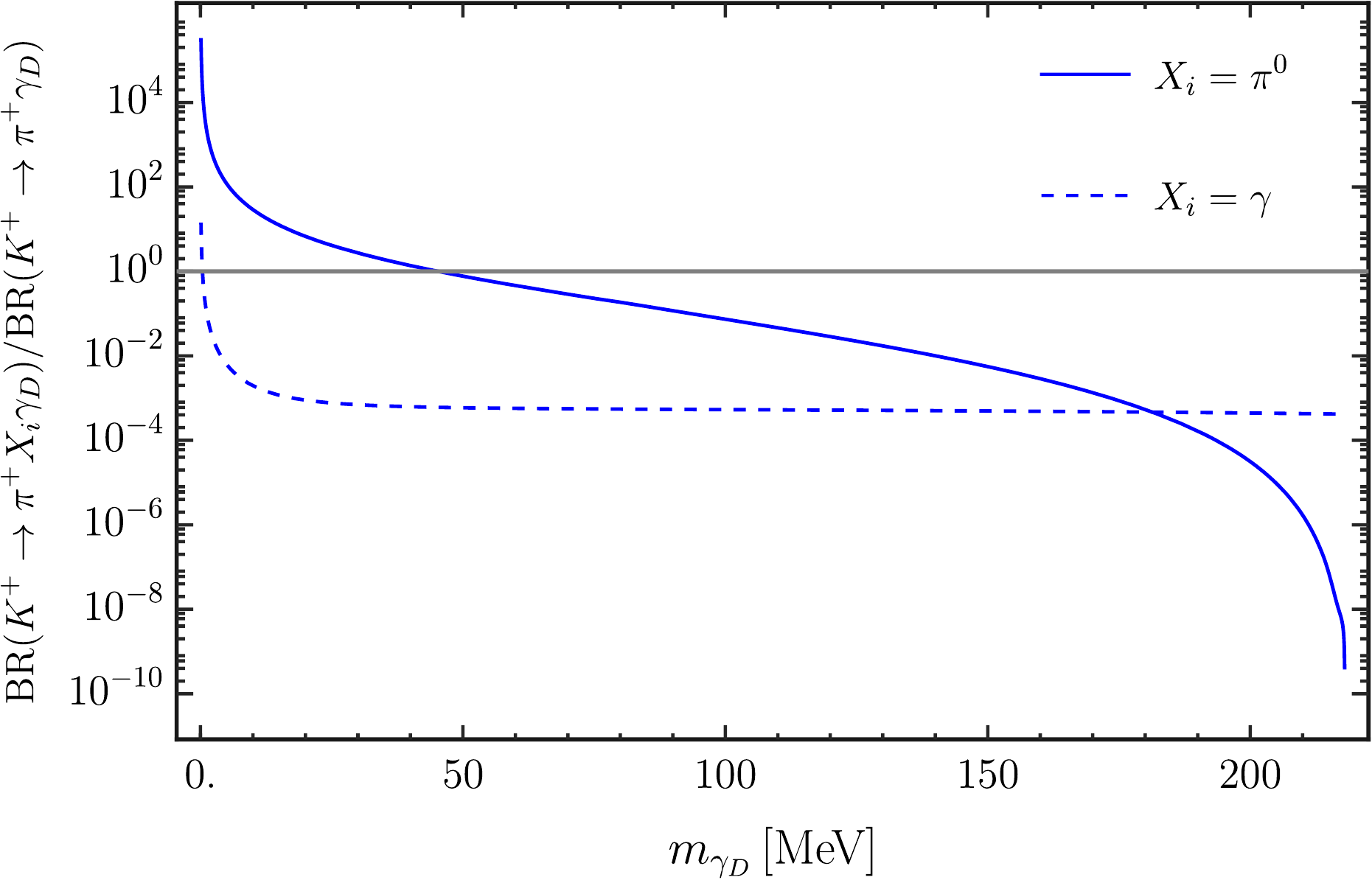}
\caption{\small
Ratio of the $\BR$s
$\BR(K^+\to\pi^+\pi^0\gammaD)$ (continuous line) and
$\BR(K^+\to\pi^+\gamma\gammaD)$ (dashed line) over $\BR(K^+\to\pi^+ \gammaD)$ as a function of the dark-photon mass $\mDP$.
}
\label{BRratios}
\end{figure}

A useful comparison of the relative importance of the two three-body
decay modes is provided in Fig.~\ref{BRratios}, where their $\BR$s are normalized to that of the non-radiative two-body decay
$K^+\to\pi^+\gammaD$. A striking feature is the very different
dependence of the two ratios on the dark-photon mass. The ratio
associated with $K^+\to\pi^+\pi^0\gammaD$ decreases rapidly with
increasing $\mDP$, as a consequence of the increasingly severe
phase-space suppression, whereas the corresponding ratio for the
radiative $K^+\to\pi^+\gamma\gammaD$ channel remains approximately
constant over a large fraction of the mass range.

This behavior can be understood, at least in part, as a consequence
of the IR enhancement of the radiative decay discussed in
Sec.~\ref{sec:Kdp2}. In the massive dark-photon case, the soft-photon
contribution is controlled by the non-radiative
$K^+\to\pi^+\gammaD$ amplitude and partially compensates for the
reduction of the available phase space as $\mDP$ increases. As a
result, the radiative one-pion mode overtakes the two-pion channel
at around $\mDP\simeq180~{\rm MeV}$ and becomes the dominant
three-body decay mode at larger dark-photon masses. This feature
highlights the complementarity of the two channels and, in particular,
the relevance of $K^+\to\pi^+\gamma\gammaD$ searches in the
higher-mass region.

An additional important feature emerging from Fig.~\ref{BRratios}
concerns the comparison with the non-radiative two-body decay
$K^+\to\pi^+\gammaD$, which corresponds experimentally to the
$K^+\to\pi^+X$ missing-mass signature conventionally used in
searches for invisible massive particles. For dark-photon production
through the FCNC magnetic-dipole interaction, the two-pion mode
$K^+\to\pi^+\pi^0\,\gammaD$ has a larger $\BR$ than
$K^+\to\pi^+\gammaD$ in the low-mass region, up to dark-photon
masses of approximately $50~{\rm MeV}$. This is a consequence of
the strong suppression of the two-body amplitude as the massless
limit is approached. Therefore, for a massive dark photon coupled
predominantly through the magnetic-dipole interaction, the
$K^+\to\pi^+\pi^0\,\gammaD$ decay can provide a particularly
sensitive discovery channel in the low-mass region below $50~{\rm MeV}$, complementary to the conventional mode of $K^+\to\pi^+X$ search.

\begin{figure*}[t!]
\centering

\includegraphics[width=0.45\textwidth]
{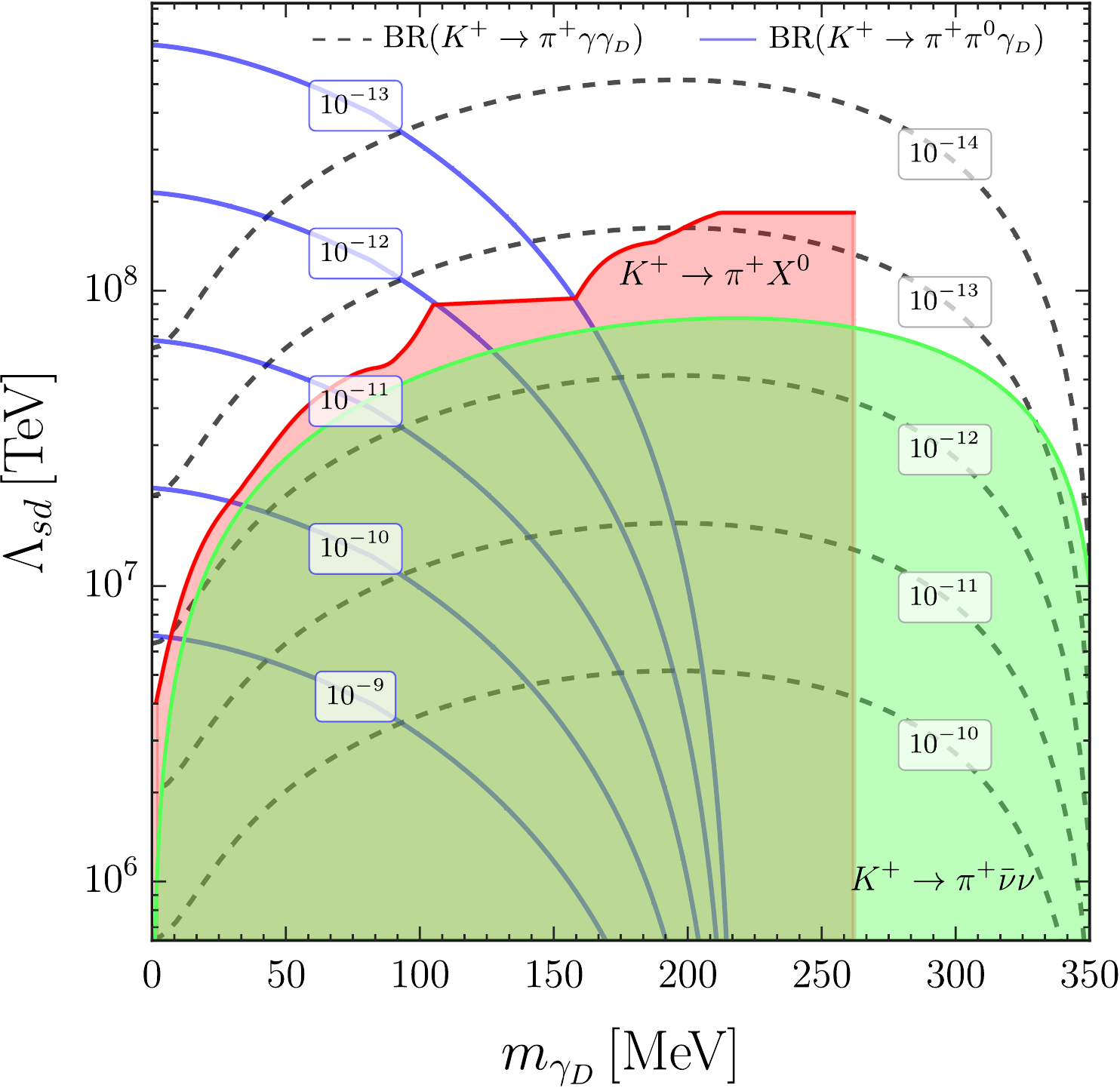}
\hspace{0.05\textwidth}
\includegraphics[width=0.45\textwidth]
{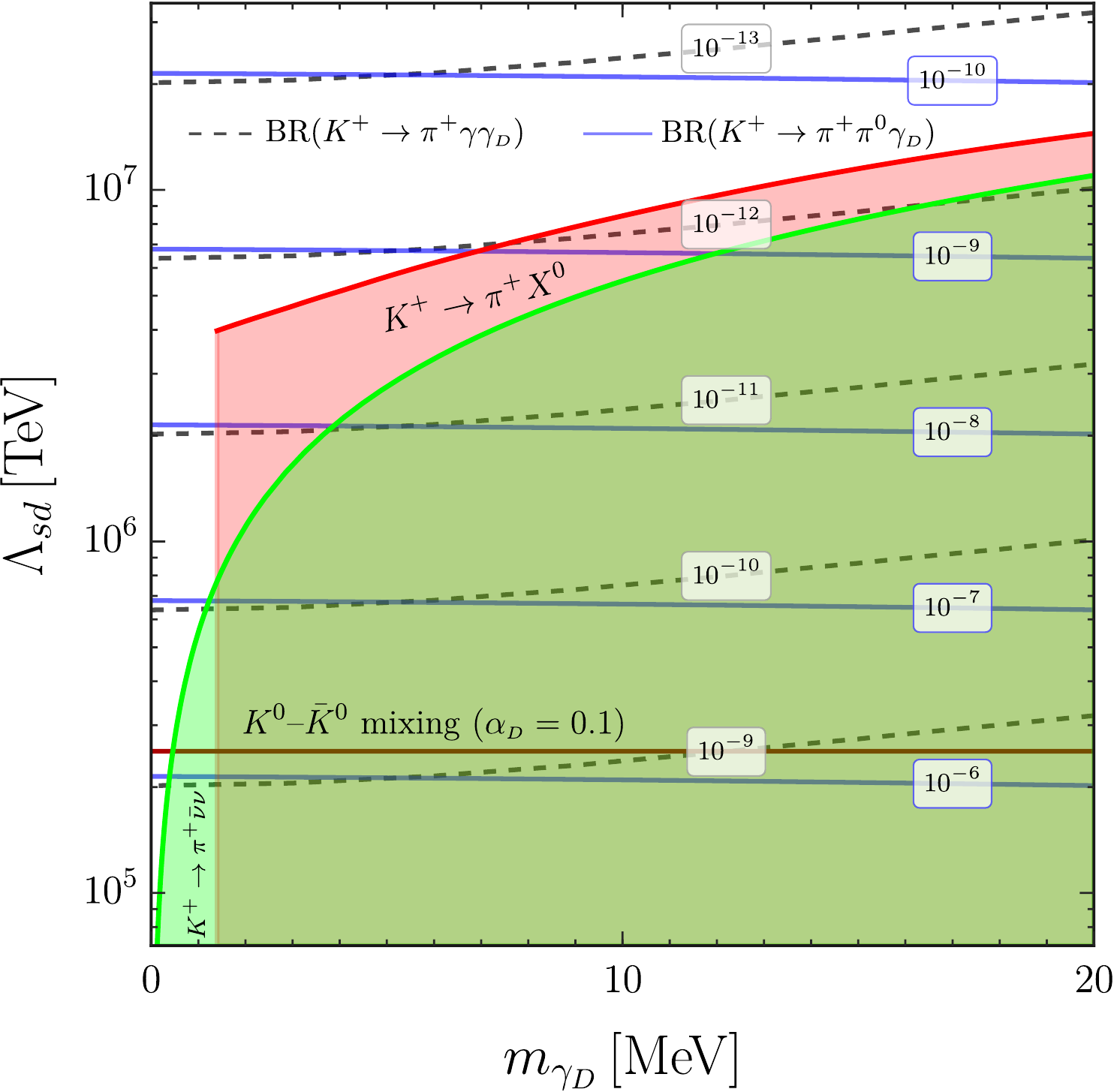}
\caption{\small Contour plots in the ($\Lambdasd,\mDP$) plane for the $\BR$s of
$K^+\to\pi^+\pi^0\,\gammaD$ (solid lines) and
$K^+\to\pi^+\gamma\,\gammaD$ (dashed lines) in the dark-photon
mass range $1\, {\rm MeV}<\mDP<350\,{\rm MeV}$ (left panel), and a zoomed-in
view of the region $1\, {\rm MeV}<\mDP<20\,{\rm MeV}$ (right  panel).
The values shown in the boxes indicate the $\BR$s
associated with each
contour line. The colored regions are excluded by the upper bound on
${\rm BR}(K^+\to\pi^+X)$ (red) at 90\% C.L.~\cite{NA62:2025upx} and by the constraint from
${\rm BR}(K^+\to\pi^+\nu\bar{\nu})$ at the $2\sigma$ level (green)~\cite{NA62:2024pjp,NA62:2026rwr}. The dark-green region corresponds to the overlap of the two excluded regions. The horizontal continuous line (brown) on the right plot
corresponds to the upper bounds on $\Lambdasd$ from \KKbar~mixing
contraints.}
\label{contourplot}
\end{figure*}

We now turn to the phenomenologically allowed regions of parameter
space. Figure~\ref{contourplot} summarizes the main results in the
$(\Lambdasd,\mDP)$ plane. The solid and dashed contours correspond to
fixed values of the $\BR$s for
$K^+\to\pi^+\pi^0\gammaD$ and $K^+\to\pi^+\gamma\gammaD$,
respectively. The upper panel covers the dark-photon mass range
$1~{\rm MeV}<\mDP<350~{\rm MeV}$, while the lower panel provides a zoom of the
low-mass region, $1~{\rm MeV} <\mDP<20~{\rm MeV}$.

The experimentally excluded regions are also displayed, showing the
impact of the bounds from $K^+\to\pi^+X$ at $90\%$ C.L.~
\cite{NA62:2025upx} and from $K^+\to\pi^+\nu\bar{\nu}$ at the
$2\sigma$ level~\cite{NA62:2024pjp,NA62:2026rwr}, together with their
overlap. In the low-mass region, the horizontal solid line indicates
the lower bound on $\Lambdasd$ obtained from \KKbar~mixing.

Several important features can be inferred from
Fig.~\ref{contourplot}. At low dark-photon masses, the
$K^+\to\pi^+\pi^0\gammaD$ channel can reach $\BR$s
substantially larger than those of $K^+\to\pi^+\gamma\gammaD$ for the
same values of $\Lambdasd$. However, its $\BR$ decreases
rapidly as $\mDP$ increases and eventually vanishes at the kinematic
endpoint. The radiative $K^+\to\pi^+\gamma\gammaD$ channel, instead,
shows a much milder dependence on $\mDP$, in agreement with the
behavior exhibited by the ratios in Fig.~\ref{BRratios}. Consequently,
although the two-pion channel provides the larger $\BR$ at
small $\mDP$, the radiative one-pion channel becomes increasingly
important at larger masses and remains kinematically accessible over
a substantially wider dark-photon mass range.

\begin{table}[h!]
\centering
\begin{tabular}{cccc}
\hline\hline
\rule{0pt}{4.0ex}
\!\!\!\!\!
$\mDP[{\rm MeV}]$ &
$\Lambdasd[{\rm TeV}]$ &
${\rm BR}\bigl(\text{\small $K^+\!\!\!\to\!\pi^+\pi^0\gammaD$}\bigr)$ &
${\rm BR}\bigl(\text{\small $K^+\!\!\!\to\!\pi^+\gamma\gammaD$}\bigr)$
\\[0.7ex]
\hline
\rule{0pt}{2.9ex}
$5$   & $5.7\times10^{6}$ & $1.41\times10^{-9}$  & $1.40\times10^{-12}$ \\
$10$  & $8.4\times10^{6}$ & $6.18\times10^{-10}$ & $7.92\times10^{-13}$ \\
$30$  & $2.0\times10^{7}$ & $9.52\times10^{-11}$ & $4.56\times10^{-13}$ \\
$50$  & $3.4\times10^{7}$ & $2.44\times10^{-11}$ & $3.53\times10^{-13}$ \\
$80$  & $5.4\times10^{7}$ & $5.71\times10^{-12}$ & $3.17\times10^{-13}$ \\
$100$ & $7.5\times10^{7}$ & $1.75\times10^{-12}$ & $2.38\times10^{-13}$ \\
$150$ & $9.3\times10^{7}$ & $1.57\times10^{-13}$ & $2.62\times10^{-13}$ \\
$200$ & $1.7\times10^{8}$ & $3.67\times10^{-16}$ & $9.69\times10^{-14}$ \\
\hline\hline
\end{tabular}
\caption{
Upper bounds on
${\rm BR}(K^+\to\pi^+\pi^0\,\gammaD)$ and
${\rm BR}(K^+\to\pi^+\gamma\,\gammaD)$ obtained from the minimal
allowed value of $\Lambdasd$, also reported in the second column,
as set by the upper bounds on
${\rm BR}(K^+\to\pi^+X)$ at 90\% C.L.~\cite{NA62:2025upx},
with $X$ denoting missing energy, for representative values of the
dark-photon mass. For the bound at $\mDP=200~{\rm MeV}$, we used the
$2\sigma$ constraint on the new-physics contribution to
${\rm BR}(K^+\to\pi^+\nu\bar{\nu})$~
\cite{NA62:2024pjp,NA62:2026rwr}.}
\label{tableBR3}
\end{table}

The corresponding upper bounds on the two $\BR$s are reported
in Table~\ref{tableBR3} for representative values of the dark-photon
mass. Two different experimental constraints determine the lower bound
on $\Lambdasd$ over the mass range considered. For
$\mDP\lesssim250~{\rm MeV}$, the dominant constraint is provided by
the NA62 search for $K^+\to\pi^+X$, with $X$ denoting an invisible
particle. At larger dark-photon masses, where this constraint is no
longer available, we require the additional contribution to
$K^+\to\pi^+\nu\bar{\nu}$ not to exceed the experimentally allowed
range at the $2\sigma$ level, after subtracting the SM contribution.

As shown in Table~\ref{tableBR3}, these constraints lead to a rapid
decrease of the maximal allowed
$\BR(K^+\to\pi^+\pi^0\gammaD)$ with increasing $\mDP$, whereas the
upper bound on $\BR(K^+\to\pi^+\gamma\gammaD)$ exhibits a considerably
milder dependence on the dark-photon mass. The latter behavior is
again consistent with the IR enhancement of the radiative
channel discussed above. Taken together, Fig.~\ref{contourplot} and
Table~\ref{tableBR3} identify the branching-ratio sensitivities and
dark-photon mass regions that are most relevant for dedicated searches
at NA62 and future high-intensity kaon facilities.

\section{Conclusions}
\label{sec:conclusions}

We have analyzed the sensitivity required to probe the two-pion
$K^+\to\pi^+\pi^0\gammaD$ and radiative one-pion
$K^+\to\pi^+\gamma\gammaD$ decays at NA62 and future kaon facilities,
considering both massless and massive dark-photon scenarios,  together with the existing two-body $K^{+}\to \pi^{+} \gamma_{D}$ constraint that sets the massive-case bounds. In the
massless case, we assume the FCNC magnetic-dipole interaction associated
with the $s\to d$ transition as the leading dark-photon production
mechanism. For a massive dark photon, we consider both production through
the magnetic-dipole operator and the standard kinetic-mixing mechanism.

For dark-photon production through the magnetic-dipole interaction, we
have taken into account the most relevant constraints on the effective
scale $\Lambdasd$. In the massless case, the strongest laboratory
constraint considered here arises from \KKbar~mixing, and
these constraints translate into the upper bounds
\bea
{\rm BR}(K^+\to\pi^+\pi^0\gammaD)&\lesssim&7\times10^{-7}\,,
\nonumber\\
{\rm BR}(K^+\to\pi^+\gamma\gammaD)&\lesssim&6\times10^{-10}\,.
\eea
These values therefore set the approximate sensitivity targets for massles dark
photon searches  at NA62 and future kaon facilities.

In the
massive case, more stringent constraints are obtained from the
$K^+\to\pi^+X$ search, with $X$ denoting missing energy, and from the measured $K^+\to\pi^+\nu\bar\nu$ $\BR$.
Then, in this case, the required sensitivity becomes stringent as the dark-photon mass increases. In particular, for
$K^+\to\pi^+\pi^0\gammaD$, branching-ratio sensitivities below
approximately $10^{-9}$ are required already for
$\mDP\gtrsim5~{\rm MeV}$. In the range
$\mDP\simeq30$--$100~{\rm MeV}$, the corresponding upper bounds decrease
from approximately ${\cal O}(10^{-10})$ to ${\cal O}(10^{-12})$.

An interesting feature emerges from the comparison of the two
three-body decay modes. While the $K^+\to\pi^+\pi^0\gammaD$ rate is
rapidly suppressed with increasing dark-photon mass because of the
reduction of the available phase space, the radiative
$K^+\to\pi^+\gamma\gammaD$ mode exhibits a much milder dependence on
$\mDP$. This behavior is partly related to the IR enhancement
associated with the soft-photon region, which compensates for part of
the phase-space suppression. As a consequence, the radiative one-pion
channel overtakes the two-pion mode at 
$\mDP\simeq180~{\rm MeV}$ and becomes the dominant three-body
dark-photon production channel induced by the FCNC magnetic-dipole
interaction at larger masses. The two decay modes therefore provide
complementary probes of the magnetic-dipole interaction over different
regions of the dark-photon mass spectrum.

Finally, we have investigated dark-photon production through kinetic
mixing in the massive scenario. Existing constraints on the kinetic
mixing parameter $\varepsilon$ in the mass range relevant to our
analysis strongly suppress both decay modes. Within the approximate
rescaling adopted here, the corresponding $\BR$s are
estimated to lie at the level of ${\cal O}(10^{-15}$--$10^{-13})$,
requiring sensitivities well beyond those presently achieved at kaon
experiments. Although a complete treatment at larger dark-photon
masses would require including finite-mass effects, in particular the
longitudinal polarization, these estimates indicate that production
through the FCNC magnetic-dipole interaction is the phenomenologically
most relevant mechanism for the kaon decay signatures investigated
here.

The large kaon samples available at NA62 make these decay modes
potentially interesting targets for dedicated searches, particularly
where the magnetic-dipole interaction allows $\BR$s approaching
the sensitivities discussed above. Although a dedicated experimental
analysis of acceptance, kinematic selection, photon-energy cuts, and
background rejection is needed to determine the actual sensitivity,
our results identify the dark-photon mass and branching-ratio regions
where such searches could be particularly motivated.

\section*{Acknowledgements}
We would like to thank Gilberto Colangelo and Evgueni Goudzovski, for useful discussions. E.G. is also grateful to the CERN Theory Division for its kind hospitality during the initial stage of this work.
This work was supported by the Estonian Research Council grants TARISTU24-TK10, TARISTU24-TK3, RVTT3, and by the CoE TK 202 ``Foundations of the Universe’', and by the CERN Science Consortium of Estonia, grant RVTT3. The work of K.M. was supported by the Estonian Research Council grant PUTJD1256

\clearpage

\onecolumngrid
\section*{Appendix}
Here we provide the expressions for the dimensionless form factors
introduced in Secs.~\ref{sec:Kdp1} and \ref{sec:Kdp2} in terms of
Passarino--Veltman functions, which can be reduced to the standard
scalar basis.

In Sec. \ref{sec:Kdp1}, the form factor $F(z_1,z_2)$ in \eq{CM} of two-pion $K^+\to \pi^+ \pi^0\, \gammaD$ decay is given by
\bea
F(z_1,z_2)&=&
\frac{2M^3 m_K^3}{\Lambdasd\,f^3\,\pi^2}
\Big[M^2\, D_0\left(\mDP^2,\mpi^2,\mpi^2,\mK^2;\mDP^2+2 \mK^2 z_1+\mpi^2,
   \mK^2 (1-2z_1-2z_2)-\mDP^2;M,M,M,M\right)\nonumber \\
   &-&D_{00}\left(\mDP^2,\mpi^2,\mpi^2,\mK^2;\mDP^2+2 \mK^2z_1+\mpi^2,
   \mK^2 (1-2z_1-2z_2)-\mDP^2;M,M,M,M\right)\nonumber\\
   &+&\left(z_1\leftrightarrow z_2\right)\, .
   \Big]
\eea

Below we provide the form factors $F_{A,B,C}(z_1,z_2)$
appearing in the amplitude for the one-pion
$K^+\to \pi^+ \gamma\, \gammaD$ decay in \eq{OM}. In  order
to simplify the notatiom, the $(z_1,z_2)$ dependence has been omitted
in the form factors.
\bea
F_A &=& \frac{ie\, M^3\mK^4}{6\,\Lambdasd\,f^2\,\pi^2}
\Bigg( \frac{1}{\mK^2}\Big[
\Delta(z_1,z_2)
+ C_0\left(0,\mDP^2,\mDP^2+2\mK^2z_2;M,M,M\right)
+ C_0\left(\mDP^2,0,\mDP^2+2\mK^2z_2;M,M,M\right)
\nonumber\\
&&
-3 C_0\left(\mDP^2,\mK^2(1-2z_1-2z_2)-\mDP^2,\mK^2;M,M,M\right)\Big]
\nonumber\\
&&
-2 z_1 D_3\left(0,\mDP^2,\mpi^2,\mK^2;
\mDP^2+2\mK^2z_2,\mDP^2+\mpi^2+2\mK^2z_1;M,M,M,M\right)
\nonumber\\
&&
-2 z_1 D_3\left(\mDP^2,0,\mpi^2,\mK^2;
\mDP^2+2\mK^2z_2,\mK^2(1-2z_1-2z_2)-\mDP^2;M,M,M,M\right)
\nonumber\\
&&
+4 z_1 D_3\left(\mDP^2,\mpi^2,0,\mK^2;
\mDP^2+\mpi^2+2\mK^2z_1,\mK^2(1-2z_1-2z_2)-\mDP^2;M,M,M,M\right)
\nonumber\\
&&
-2 z_1 D_2\left(0,\mDP^2,\mpi^2,\mK^2;
\mDP^2+2\mK^2z_2,\mDP^2+\mpi^2+2\mK^2z_1;M,M,M,M\right)
\nonumber\\
&&
-2 z_1 D_2\left(\mDP^2,0,\mpi^2,\mK^2;
\mDP^2+2\mK^2z_2,\mK^2(1-2z_1-2z_2)-\mDP^2;M,M,M,M\right)
\nonumber\\
&&
+4 z_1 D_2\left(\mDP^2,\mpi^2,0,\mK^2;
\mDP^2+\mpi^2+2\mK^2z_1,\mK^2(1-2z_1-2z_2)-\mDP^2;M,M,M,M\right)
\nonumber\\
&&
-2 z_1 D_1\left(\mDP^2,0,\mpi^2,\mK^2;
\mDP^2+2\mK^2z_2,\mK^2(1-2z_1-2z_2)-\mDP^2;M,M,M,M\right)
\nonumber\\
&&
+4 z_1 D_1\left(\mDP^2,\mpi^2,0,\mK^2;
\mDP^2+\mpi^2+2\mK^2z_1,\mK^2(1-2z_1-2z_2)-\mDP^2;M,M,M,M\right)
\nonumber\\
&&
+ \left(\rDP-1+2 z_2\right)
D_0\left(\mDP^2,0,\mpi^2,\mK^2;\mDP^2+2\mK^2z_2,
\mK^2(1-2z_1-2z_2)-\mDP^2;M,M,M,M\right)
\nonumber\\
&&
-\left(\rp+2 z_1\right)
D_0\left(0,\mDP^2,\mpi^2,\mK^2;\mDP^2+2\mK^2z_2,
\mDP^2+\mpi^2+2\mK^2z_1;M,M,M,M\right)
\nonumber\\
&&
+\left(2\rp + 4z_1\right)
D_0\left(\mDP^2,\mpi^2,0,\mK^2;\mDP^2+\mpi^2+2\mK^2z_1,
\mK^2(1-2z_1-2z_2)-\mDP^2;M,M,M,M\right)
\Bigg),
\label{FA}
\\
F_B &=& \frac{ie\, M^3 \mK^4}{12\,\Lambdasd\, f^2\, z_2\, \pi^2}
\Bigg\{ \frac{1}{\mK^2}\Big[
\frac{12\rDP}{R-2z_2}
C_0\left(
\mDP^2,\mpi^2,\mDP^2+\mpi^2+2\mK^2z_1;
M,M,M\right)
\nonumber\\
&&
-\frac{12\rDP}
{R}
C_0\left(
\mDP^2,\mK^2(1-2z_1-2z_2)-\mDP^2,\mK^2;
M,M,M\right)\Big]
\nonumber\\
&&
+8 \rDP
D_0\left(
\mDP^2,\mpi^2,0,\mK^2;
\mDP^2+\mpi^2+2\mK^2z_1,
\mK^2(1-2z_1-2z_2)-\mDP^2;
M,M,M,M\right)
\nonumber\\
&&
-4 (\rDP+z_2) 
D_3\left(
0,\mDP^2,\mpi^2,\mK^2;
\mDP^2+2\mK^2z_2,
\mDP^2+\mpi^2+2\mK^2z_1;
M,M,M,M\right)
\nonumber\\
&&
-4 (\rDP+z_2)
D_3\left(
\mDP^2,0,\mpi^2,\mK^2;
\mDP^2+2\mK^2z_2,
\mK^2(1-2z_1-2z_2)-\mDP^2;
M,M,M,M\right)
\nonumber\\
&&
+8 (\rDP+z_2)
D_3\left(
\mDP^2,\mpi^2,0,\mK^2;
\mDP^2+\mpi^2+2\mK^2z_1,
\mK^2(1-2z_1-2z_2)-\mDP^2;
M,M,M,M\right)
\nonumber\\
&&
+8 (\rDP+z_2)
D_2\left(
\mDP^2,\mpi^2,0,\mK^2;
\mDP^2+\mpi^2+2\mK^2z_1,
\mK^2(1-2z_1-2z_2)-\mDP^2;
M,M,M,M\right)
\Bigg\},
\label{FB}
\eea
\bea
F_C &=& -\frac{i e\, M^3\mK^4}{12\,\Lambdasd\,f^2\,\pi^2}
\Bigg\{\frac{1}{\mK^2}\Big[
-\frac{12 z_1}{R-2z_2}
C_0\left(
\mDP^2,\mpi^2,\mDP^2+\mpi^2+2\mK^2z_1;
M,M,M\right)
\nonumber\\
&&
+6 C_0\left(
\mDP^2,\mK^2(1-2z_1-2z_2)-\mDP^2,\mK^2;
M,M,M\right)\Big]
\nonumber\\
&&
+ R
D_0\left(
0,\mDP^2,\mpi^2,\mK^2;
\mDP^2+2\mK^2z_2,
\mDP^2+\mpi^2+2\mK^2z_1;
M,M,M,M\right)
\nonumber\\
&&
- (R-4 z_1)
D_0\left(
\mDP^2,0,\mpi^2,\mK^2;
\mDP^2+2\mK^2z_2,
\mK^2(1-2z_1-2z_2)-\mDP^2;
M,M,M,M\right)
\nonumber\\
&&
-2
\left(R+4 z_1\right)
D_0\left(
\mDP^2,\mpi^2,0,\mK^2;
\mDP^2+\mpi^2+2\mK^2z_1,
\mK^2(1-2z_1-2z_2)-\mDP^2;
M,M,M,M\right)
\nonumber\\
&&
+4 z_1
\Big[
D_3\left(
0,\mDP^2,\mpi^2,\mK^2;
\mDP^2+2\mK^2z_2,
\mDP^2+\mpi^2+2\mK^2z_1;
M,M,M,M\right)
\nonumber\\
&&\hspace{0.7cm}
+D_3\left(
\mDP^2,0,\mpi^2,\mK^2;
\mDP^2+2\mK^2z_2,
\mK^2(1-2z_1-2z_2)-\mDP^2;
M,M,M,M\right)
\nonumber\\
&&\hspace{0.7cm}
-2D_3\left(
\mDP^2,\mpi^2,0,\mK^2;
\mDP^2+\mpi^2+2\mK^2z_1,
\mK^2(1-2z_1-2z_2)-\mDP^2;
M,M,M,M\right)
\nonumber\\
&&\hspace{0.7cm}
-2 D_2\left(
\mDP^2,\mpi^2,0,\mK^2;
\mDP^2+\mpi^2+2\mK^2z_1,
\mK^2(1-2z_1-2z_2)-\mDP^2;
M,M,M,M\right)
\nonumber\\
&&\hspace{0.7cm}
+D_2\left(
0,\mDP^2,\mpi^2,\mK^2;
\mDP^2+2\mK^2z_2,
\mDP^2+\mpi^2+2\mK^2z_1;
M,M,M,M\right)
\nonumber\\
&&\hspace{0.7cm}
+D_2\left(
\mDP^2,0,\mpi^2,\mK^2;
\mDP^2+2\mK^2z_2,
\mK^2(1-2z_1-2z_2)-\mDP^2;
M,M,M,M\right)
\nonumber\\
&&\hspace{0.7cm}
+ D_1\left(
\mDP^2,0,\mpi^2,\mK^2;
\mDP^2+2\mK^2z_2,
\mK^2(1-2z_1-2z_2)-\mDP^2;
M,M,M,M\right)
\nonumber\\
&&\hspace{0.61cm}
-2D_1\left(
\mDP^2,\mpi^2,0,\mK^2;
\mDP^2+\mpi^2+2\mK^2z_1,
\mK^2(1-2z_1-2z_2)-\mDP^2;
M,M,M,M\right)
\Big]
\Bigg\},
\label{FC}
\eea

where $e$ is the unit of electric charge, $R$ is defined below \eq{fABC}, and the $\Delta(z_1,z_2)$ function is given by
\bea
\Delta(z_1,z_2)&=&-\frac{1}{4\,M^2\,\mK^4 z_2^2}
\Bigg\{
2\mK^2z_2(\mK^2z_2-\mDP^2)
+\mDP^2(\mDP^2+\mK^2z_2)\DiscB(\mDP^2,M,M)
\nonumber\\
&&\quad
-\left(\mDP^4+\mDP^2\mK^2z_2-2\mK^4z_2^2\right)
\DiscB(\mDP^2+2\mK^2z_2,M,M)
\nonumber\\
&&\quad
+2\mK^4z_2^2\log\!\left(\frac{\mu^2}{M^2}\right)
+M^2(\mDP^2-\mK^2z_2)
\log^2\!\left[
\frac{2M^2-\mDP^2+
\sqrt{\mDP^2(\mDP^2-4M^2)}}{2M^2}
\right]
\nonumber\\
&&\quad
-M^2(\mDP^2-\mK^2z_2)
\log^2\!\left[
\frac{2M^2-\mDP^2-2\mK^2z_2+
\sqrt{(\mDP^2+2\mK^2z_2)
(\mDP^2+2\mK^2z_2-4M^2)}}{2M^2}
\right]
\Bigg\}\, ,
\nonumber
\eea
where $\DiscB(s,M,M)\equiv \sqrt{1-\frac{4M^2}{s}}\,\log\!\left(
\frac{2M^2-s+\sqrt{s(s-4M^2)}}{2M^2}\right)$.
Here, the Passarino--Veltman notation for the $C_0$, $D_0$, $D_i$
($i=1,2,3$), and $D_{00}$ functions is that introduced in Sec.\ref{sec:Kdp1}.


\bibliographystyle{utphys}
\bibliography{K-dark-photons.bib}

\end{document}